\documentclass[10pt,twocolumn,showpacs,preprintnumbers,amsmath,amssymb,aps,prb,superscriptaddress,floatfix,footinbib]{revtex4-1}

\usepackage{graphicx}
\usepackage{dcolumn}
\usepackage{lipsum}
\usepackage{bm}
\usepackage{epstopdf}
\usepackage{amsmath}
\usepackage{amssymb}
\usepackage{hyperref}

\begin{document}



\title{Magnetic-field-induced FM-AFM metamagnetic transition and strong
  negative magnetoresistance in Mn$_{1/4}$NbS$_2$ under pressure.}
 \author{S.\ Polesya} 
 \affiliation{%
   Department  Chemie,  Physikalische  Chemie,  Universit\"at  M\"unchen,
   Butenandstr.\ 5-13, 81377  M\"unchen, Germany\\}
 \author{S.\ Mankovsky}
 \affiliation{%
   Department  Chemie,  Physikalische  Chemie,  Universit\"at  M\"unchen,
   Butenandstr.\  5-13, 81377 M\"unchen, Germany\\}
\author{P. G. Naumov} 
\affiliation{%
Max Planck Institute for Chemical Physics of Solids, N\"othnitzer
Str. 40, D-01187 Dresden, Germany\\}
\affiliation{%
Shubnikov Institute of Crystallography, Russian Academy of Sciences, Moscow 119333, Russia\\
}
\author{M. A. ElGhazali} 
\affiliation{%
Max Planck Institute for Chemical Physics of Solids, N\"othnitzer Str. 40,
D-01187 Dresden, Germany\\}
\author{W. Schnelle} 
\affiliation{%
Max Planck Institute for Chemical Physics of Solids, N\"othnitzer Str. 40,
D-01187 Dresden, Germany\\}
\author{S.\ Medvedev } 
\affiliation{%
Max Planck Institute for Chemical Physics of Solids, N\"othnitzer Str. 40,
D-01187 Dresden, Germany\\}
\author{S.\ Mangelsen} 
\affiliation{%
Inst.\ f\"ur Anorgan.\ Chemie,  Universit\"at Kiel,
Olshausenstr.\ 40,  24098 Kiel, Germany\\}
\author{W.\ Bensch} 
\affiliation{%
Inst.\ f\"ur Anorgan.\ Chemie,  Universit\"at Kiel,
Olshausenstr.\ 40,  24098 Kiel, Germany\\}
\author{H.\ Ebert} 
 \affiliation{%
   Department  Chemie,  Physikalische  Chemie,  Universit\"at  M\"unchen,
   Butenandstr.\ 5-13, 81377  M\"unchen, Germany\\}

\date{\today}
             
\begin{abstract}

Transition metal dichalcogenides (TMDC) stand out with their high
chemical stability and the possibility to incorporate a wide range of
magnetic species between the layers. The behavior of conduction
electrons in such materials intercalated by 3d-elements is closely
related to their magnetic properties and can be sensitively controlled by
external magnetic fields. Here, we study the
magnetotransport properties of NbS$_2$ intercalated with Mn, Mn$_{1/4}$NbS$_2$,
demonstrating a complex behavior of the magnetoresistance and of the
ordinary and anomalous Hall resistivities. Application of pressure as
tuning parameter leads to the drastic changes of the magnetotransport
properties of Mn$_{1/4}$NbS$_2$ exhibiting large negative magnetoresistance up
to $–65 \%$ at 7.1 GPa. First-principles electronic structure
calculations indicates pressure-induced transition from ferromagnetic to
antiferromagnetic state. Theoretical calculations accounting for the
finite temperature magnetic properties of  Mn$_{1/4}$NbS$_2$ suggest a
field-induced metamagnetic ferromagnetic-antiferromagnetic transition as
an origin of the large negative magentoresistance. These results inspire
the development of materials for spintronic applications based on
intercalated TMDC with a well controllable metamagnetic transition.  
\end{abstract}

\pacs{Valid PACS appear here}
\maketitle

\section{Introduction}
  The transition metal dichalcogenides (TMDC) are in focus of various
 investigations since many years as they exhibit a broad spectrum of
 structure and composition dependent physical properties. 
 Being non-magnetic itself, some of the known TMDC materials 
 allow intercalation by magnetic 3d elements \cite{FY87}, having a 
 tendency to the formation of ordered compounds when the concentration of 
 intercalating elements is close to $25 \%$ or $33 \%$.
 This creates a family of TMDC-based magnetic compounds exhibiting rather 
 peculiar magnetic and transport properties, which however have not
 been well investigated so far.
 Among the properties of the intercalated TMDC's one can mention the
 high magnetic anisotropy \cite{CLM+08,CKA+09} and anomalous Hall effect
 (AHE) \cite{CLM+08,DZB+89} in Fe$_{1/4}$TaS$_2$ observed experimentally
 and discussed recently on the basis of the first-principles
 calculations \cite{KKK+11,MCK+15}. 
  A large magnetoresistance (MR) has been found in disordered
  Fe$_{x}$TaS$_2$ single crystals, up to $60\%$ at $x=0.28$
  \cite{HCM+15} and  $140\%$ at $x=0.297$ \cite{CCZM16}, that was
  attributed to the spin disorder and strong spin-orbit coupling in the
  system. A long-period helimagnetic (HM) structure along the hexagonal
  c-axis has been observed experimentally in Cr$_{1/3}$NbS$_2$,
  with the Cr magnetic moments aligned within the basal plane perpendicular to
  the c-axis \cite{TKT+12, BGT+15}. Stabilization of the HM structure in
  Cr$_{1/3}$NbS$_2$ has been confirmed theoretically by means of first
  principles calculations \cite{MPEB16}. 
  Similar to Cr$_{1/3}$NbS$_2$ helimagnetic properties have been also predicted
  for Mn$_{1/3}$NbS$_2$\cite{MPEB16}, while for Fe$_{1/3}$NbS$_2$ stability of the
  magnetic structure referred to as an ordering of the third kind  was
  demonstrated \cite{MPEB16} in agreement with experimental results
  \cite{LRI71, GBR+81, YMT+04}. The calculations demonstrate a
  transition to the AFM state also for Co$_{1/3}$NbS$_2$ and
  Ni$_{1/3}$NbS$_2$ \cite{PME19}, again in line with experiment 
  \cite{PF80}. 

  Application of pressure provides an appealing degree of freedom to manipulate
  the physical properties of solids, albeit only a few studies on the 
  3d-intercalated TMDC materials have been done so far. The pressure induced electronic
  structure modification can have either a direct impact on the physical
  properties or can lead first of all to a crystal structure transformations
  (see e.g. \cite{Ehm_Pressure_2007,  Mito_Investigation_2015}).
  As a direct impact one can mention the pressure induced changes
  of the magnetic and transport properties of Co$_{1/3}$NbS$_2$ 
  \cite{Barisic_High_2011} as well as an enhancement of the superconducting  
  properties of Fe$_{x}$NbS$_2$ \cite{Krishnan_Pressure_2018} reported
  recently. On the other hand, a suppression of the helimagnetic
  structure has been observed for Cr$_{1/3}$NbS$_2$ at $\sim 3-4$ GPa as a
  result of the structural transformations in the system \cite{Mito_Investigation_2015} .

  In the present work we focus on the magnetic properties of Mn intercalated
  2H-NbS$_2$ under pressure, with $25 \%$ intercalation
  concentration. In the first part we discuss the experimental results
  based on the magnetotransport measurements (magnetoresistance and Hall
  effect), as they are crucial for inferring information about the
  interactions between itinerant charge carriers and the magnetic
  degrees of freedom in a variety of magnetic materials. In order to
  interpret the details of the experimental measurements, theoretical
  results are discussed in the second part, which are based on DFT
  calculations and Monte Carlo simulations.

\section{Technical details}

\subsection{Experimental details} \label{SEC:Experimental-detailes}
\subsubsection{Sample preparation}
A mixture of 1 g in total of the elements (Mn, 99.95~\%, Alfa Aesar; Nb,
99.9~\% chempur; S, 99.9995~\%, Alfa Aesar) with nominal stoichiometry
Mn$_{0.265}$NbS$_{2}$ were ground and loaded into
an argon flushed silica ampoule. CBr$_{4}$ ($\sim$ 25 mg,
98~\%, Fluka) was added as transport agent. Due to the volatility of
CBr$_{4}$ the lower end of the filled ampoule was cooled with
liquid nitrogen prior to evacuation ($p \leq$
1*10$^{-4}$~mbar), after which the ampoule was sealed. 
The ampoule was placed in the natural gradient of a single tube furnace in inverse
position for the pre reaction and cleaning transport. After heating to
450$^{\circ}$C during 6~h, the temperature was
maintained there for 10~h before it was raised to
900$^{\circ}$C. After two days the ampoule was placed
in a gradient 900$^{\circ}$C $\rightarrow$ $\sim$~800$^{\circ}$C
and within 14~d crystals of several mm
diameter in the basal plane grew. The ampoule was post-annealed by
slow cooling from 800$^{\circ}$C to room temperature
within 24~h to allow for a good order of the Mn-ions which become
disordered above $\sim$
400$^{\circ}$C.\cite{Boswell_magnetic_1978,Kuroiwa_Phase_1997,Kuroiwa_Structural_2000}
After opening the ampoules the crystals were washed with water, a dilute
solution of Na$_{2}$S$_{2}$O$_{3}$
(97~\%, Gr\"ussing), water and acetone. The crystals have dimensions up
to several mm in diameter and 1~mm in height and show a metallic silver
luster. They were stored in an evacuated desiccator until usage. 

\subsubsection{Characterisation}
EDX-spectra were acquired on a Philips XL30 ESEM equipped with an EDAX
EDX-detector operated at 20~kV acceleration voltage. At least two
crystals from each batch were analysed, on each crystal the composition
was measured on three spots with an integration time of 120~s each.  

The series of 00l-reflections were measured on a Panalytical X'Pert
Pro-MPD (Cu-K$_\alpha$ radiation, $1/16^{\circ}$
divergence mask, G\"obel mirror, primary soller slit (0.04~rad) on the
incident beam path, parallel plate collimator and PIXcel 1D detector on
the diffracted beam path) in Bragg-Brentano geometry. For sample
alignment and rocking curves a receiving slit was used. 

For collecting diffraction patterns in [001]-orientation a thin slice
($\le$ 100 $\mu$m) was cut from a crystal with a thin razor blade. The
crystal piece was mounted an adhesive tape and measured in transmission
geometry on a Panalytical Empyrean diffractometer
(Cu-K$_\alpha$ radiation, $1/2^{\circ}$ divergence mask,
focussing mirror, primary soller slit (0.04~rad) on the incident beam
path, secondary soller slit (0.04~rad) and PIXcel 1D detector on the
diffracted beam path). The instrumental broadening was derived from a
measurement of LaB$_{6}$ (NIST SRM 660c) via fitting the
profile to a Thompson-Cox-Hastings Pseudo-Voigt profile with a
Pawley-fit.  

All fittings of the lattice parameters were carried out using
Topas-Academic Version 6.0\cite{Coelho_TOPAS_2018} via Pawley-fits.

\subsubsection{Ambient pressure experiments}
The magnetization was measured with a MPMS3
(SQUID-VSM, Quantum Design) on a single crystal
($m = 1.859$~mg; flat platelet showing hexagon faces) with
the magnetic field applied along the $[1000]$ (in-plane (IP))
or along the $[0001]$ (out-of-plane (OP)) direction,
respectively. $M(B)$ isotherms at selected temperatures were recorded
after cooling in zero field from $T = 200$~K, temperature sweeps were
taken during cooling in low fields. 
The heat capacity was determined on the same crystal
with the HC option in a PPMS-9 (Quantum Design) at
zero field and with $B = 9$~T applied along $[0001]$ (OP).

Transverse magnetoresistance (TMR) and Hall resistance $\rho_H$ data as
function of magnetic field along $[0001]$  
(OP) were measured in a conventional four-wire and a
five-wire (with external potentiometer) configuration, respectively. The
electrical transport option of a PPMS-9 
was used to take data in magnetic field sweeps at selected
constant temperatures. Symmetrization of the TMR and
antisymmetrization of the Hall data with respect to applied field was
performed.

\subsubsection{High pressure experiments}
For high-pressure experiments, a diamond anvil cell manufactured from
the nonmagnetic alloy MP35N and equipped with Boehler-Almax design
diamond anvils with 500-$\mu$m culets was used. The tungsten gasket was
insulated with a cubic BN/epoxy mixture. A single crystal sample of
suitable size ($120~\mu$m$ \times 120~\mu$m$ \times 10~\mu$m) was cut
and placed into the central hole of the gasket filled with NaCl as a
pressure-transmitting 
medium along with a ruby chip for pressure calibration. The electrical
leads were fabricated from 5~$\mu$m thick Pt foil and attached to the sample
in a van der Pauw configuration. Electrical resistivity was measured at
different pressures in temperature range 1.8-300~K in magnetic field up
to 9~T with Physical Property Measurement System (PPMS-9, Quantum
Design).

High pressure Raman spectra were recorded at room temperature using a
customary confocal micro-Raman spectrometer with a HeNe-laser as the
excitation source and a single-grating spectrograph with 1 cm$^{-1}$
resolution. Pressure was measured with accuracy of $\sim 0.1$ GPa using
the ruby luminescence method.


\subsection{Computational details}
\label{SEC:Computational-scheme}

The first-principles electronic structure calculations have been
performed using the spin-polarized relativistic KKR (SPR-KKR) Green function
method  \cite{SPR-KKR7.7,EKM11}.  For the angular momentum expansion of
the Green function a cutoff of $l_{max} = 3$ was applied.
All calculations have been performed within the framework of the local
spin density approximation (LSDA) to density 
functional theory (DFT) as well as beyond the level of LSDA by
accounting for correlation effects by means of the LDA+U scheme
\cite{SLP99,YAF03}. The LSDA calculations used a parametrization
for the exchange and  correlation potential as given by Vosko {\em et
  al.} \cite{VWN80}. In the LDA+U calculations the so-called 
atomic limit expression was used for double-counting correction in the
LDA+U functional, with the parameters $U = 3$~eV and $J = 0.7$~eV.

In order to investigate the equilibrium magnetic structure as well as
finite temperature magnetic properties of the compounds under
consideration, Monte Carlo simulations have been performed, which are
based on the Heisenberg model
with the exchange coupling parameters $J_{ij}$ calculated from first
principles \cite{EM09a}.

The temperature-dependent behavior of electronic resistivity of the
systems under consideration was investigated on the basis of the
Kubo-St\v{r}eda formalism in combination with the alloy analogy model.
It allows to account for thermal lattice vibrations \cite{KCME13} as
well as spin fluctuations \cite{EMC+15} treating them within the
adiabatic approximation \cite{AKH+96}.

The properties corresponding to ambient pressure have been calculated
using the structure parameters obtained in experiment.
The Mn intercalated TMDC system under investigation, Mn$_{1/4}$NbS$_2$,
exhibits a $2 \times 2$ superstructure in the Mn layers arranged
within the so-called van der Waals gap, leading to
a well defined ordered compound crystallizing in the space group
P$6_3/mmc$ (SG194), with $a = 6.67 \AA, c=12.49 \AA$. 
This implies an occupation of the $(2a)$ Wyckoff positions by Mn atoms,
and occupation of the $(2b)$ and $(6h)$ (with $x=0.5$) positions by Nb
atoms, and  $(4f)$- ($z = 0.121$) and  $(12k)$-positions ($x = 5/6,
z=0,37$) -- by S atoms.

In the investigations of the pressure dependent properties, the pressure
dependent structure parameters of Mn$_{1/4}$NbS$_2$ have not been
measured. Therefore, auxiliary calculations have been performed using
the VASP package \cite{KH93,KH94} in order to determine the relationship
between the applied pressure and lattice parameters. 
In these calculations using the GGA density functional for the exchange
and correlation potential the  PBE-parametrization 
scheme has been used as given by Perdew {\em et al.} \cite{PBE96}. 
As the van der Waals interactions may be important to describe correctly
the pressure dependent behaviour of TMDC-based system \cite{MNB+17},
these interactions have been taken into account using the DFT-D3
method for the dispersion corrections as given by Grimme et
al. \cite{GAEK10}. The Monkhorst-Pack ($8 \times 8 \times 8$) $k$-point
grid was used for the integration  over the Brillouin zone. A plane wave
basis set up to a cutoff energy of 440 eV was used for the wave function
representation.

%
\section{Results}
\subsection{Experiment}

\begin{figure}
	\centering
	\includegraphics[width=1\linewidth]{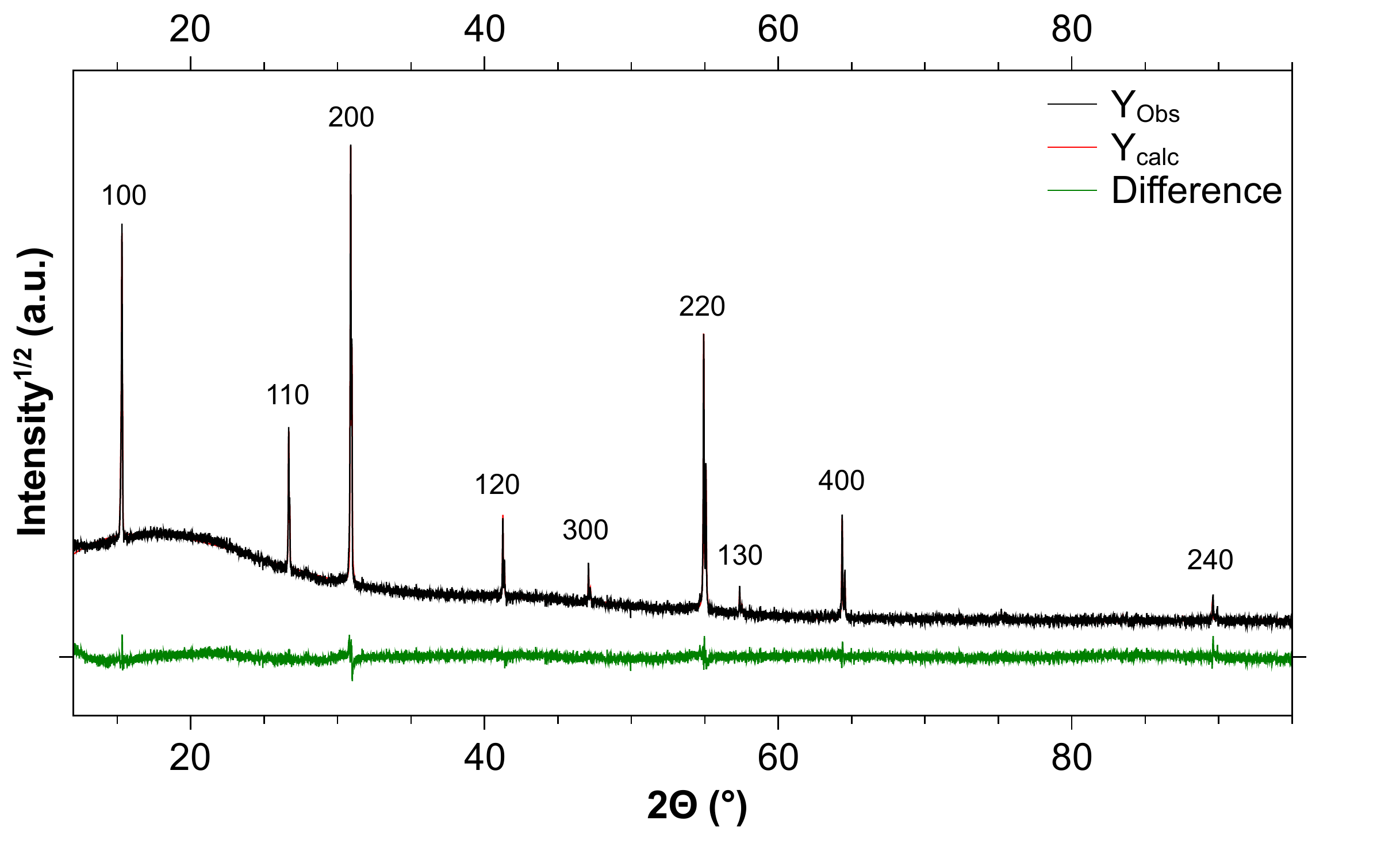}\;(a)
	\includegraphics[width=1\linewidth]{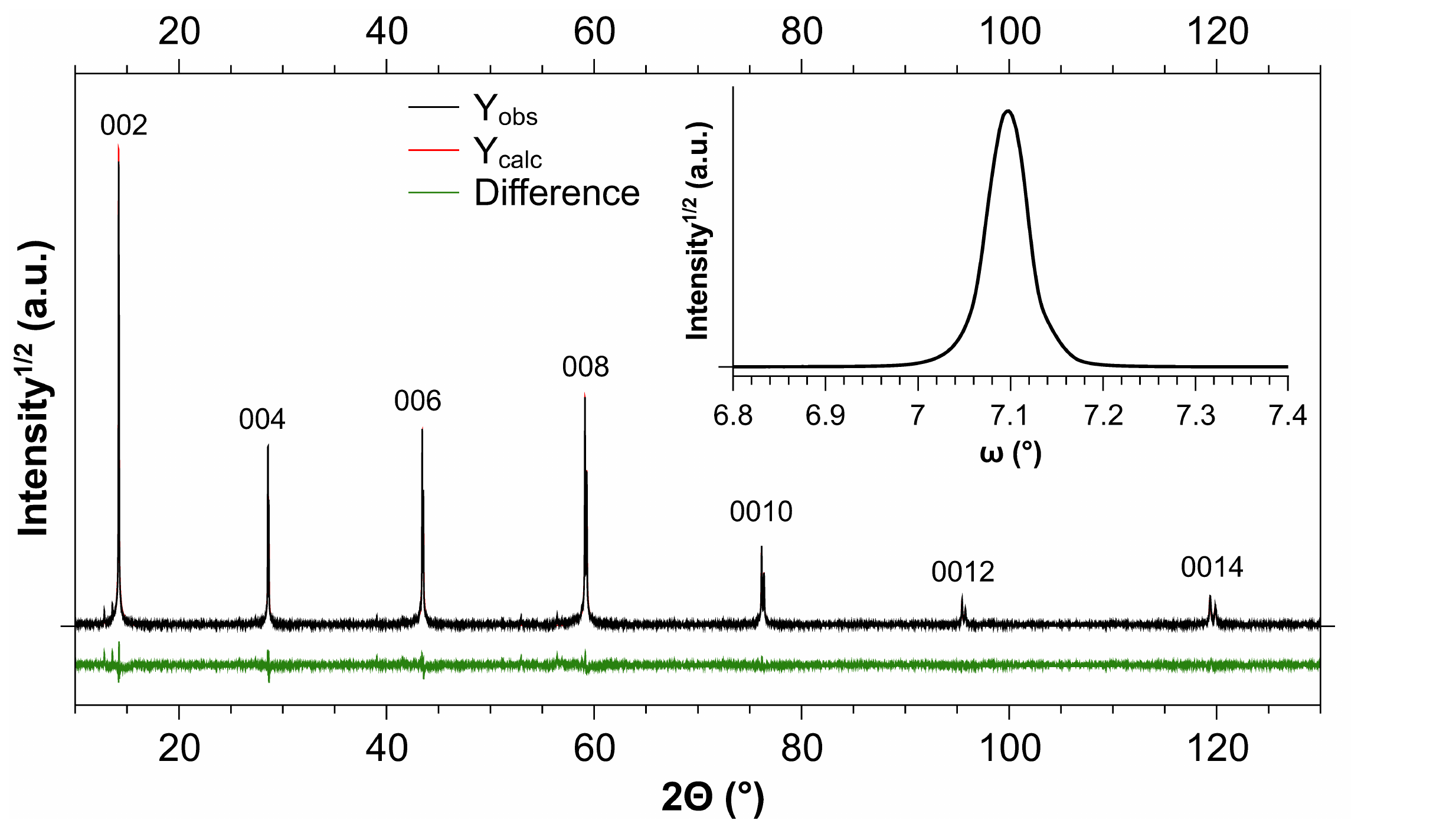}\;(b)
	\caption{Observed and calculated diffraction patterns for a sample of Mn$_{0.25}$NbS$_{2}$ for the set of hk0-reflections (a) and 00l-reflections (b). In the inset of (b) a representative rocking curve of the 002 reflection is presented.}
	\label{fig:XRD}
\end{figure}

The details of sample preparation are described in section
\ref{SEC:Experimental-detailes}. 
The elemental composition was determined by means of EDX to be
Mn$_{0.25}$NbS$_{2}$ within the limits of
experimental accuracy. The Mn ions are known to form a superstructure
of $2a \times 2a$ with respect to the 2H-NbS$_2$
host lattice. In fact, the composition determined is very close to the ideal
value for this type of superstructure. To check this, X-ray diffraction
was carried out on a thin piece cut from a crystal, which was oriented
with reflections of the [001] zone axis being under diffraction
condition.  The corresponding diffraction pattern is presented in
Fig. \ref{fig:XRD} (a).
%
There are only reflections of type \textit{h00}
and \textit{hk0} visible which have a very narrow full width at half
maximum (FWHM), indicating large coherently diffracting domains.
This holds true both for the host structure (110 reflection) as well as for
the domains of the superstructure (100 reflection). To further ensure
the crystal quality the series of 00l-reflections was measured (see
Fig. \ref{fig:XRD} (b)).
Also here a very low FWHM can be observed, indicating a large domain
size along the \textit{c}-axis.
The rocking curves measured on the 00l-reflections are of a very low
average FWHM ($0.06^\circ$, an example is shown in the inset of
Fig. \ref{fig:XRD} (b)), underscoring the high quality of the crystals.

From the XRD data
presented here we can conclude that only the $2a \times 2a$ 
type superstructure is formed by the Mn-ions with long range order. This
is evidenced by the presence of only one set of 00l-reflections and in
particular by the diffraction pattern of the [001] zone axis, where only
reflections belonging to this type of superstructure are present.
The lattice parameters are $a = b = 6.6715(4)~\AA$ and \textit{c} =
12.493163(3)~$\AA$, in very good agreement with reports from
literature.\cite{Onuki_Magnetic_1986, LRI71,Blanc-Soreau1976} These 
measurements showed reproducible results on several crystals from the
same batch. In summary, the samples can be described as nearly perfect
Mn$_{1/4}$NbS$_2$.
      
\begin{figure}[h]
\includegraphics[width=0.4\textwidth,angle=0,clip]{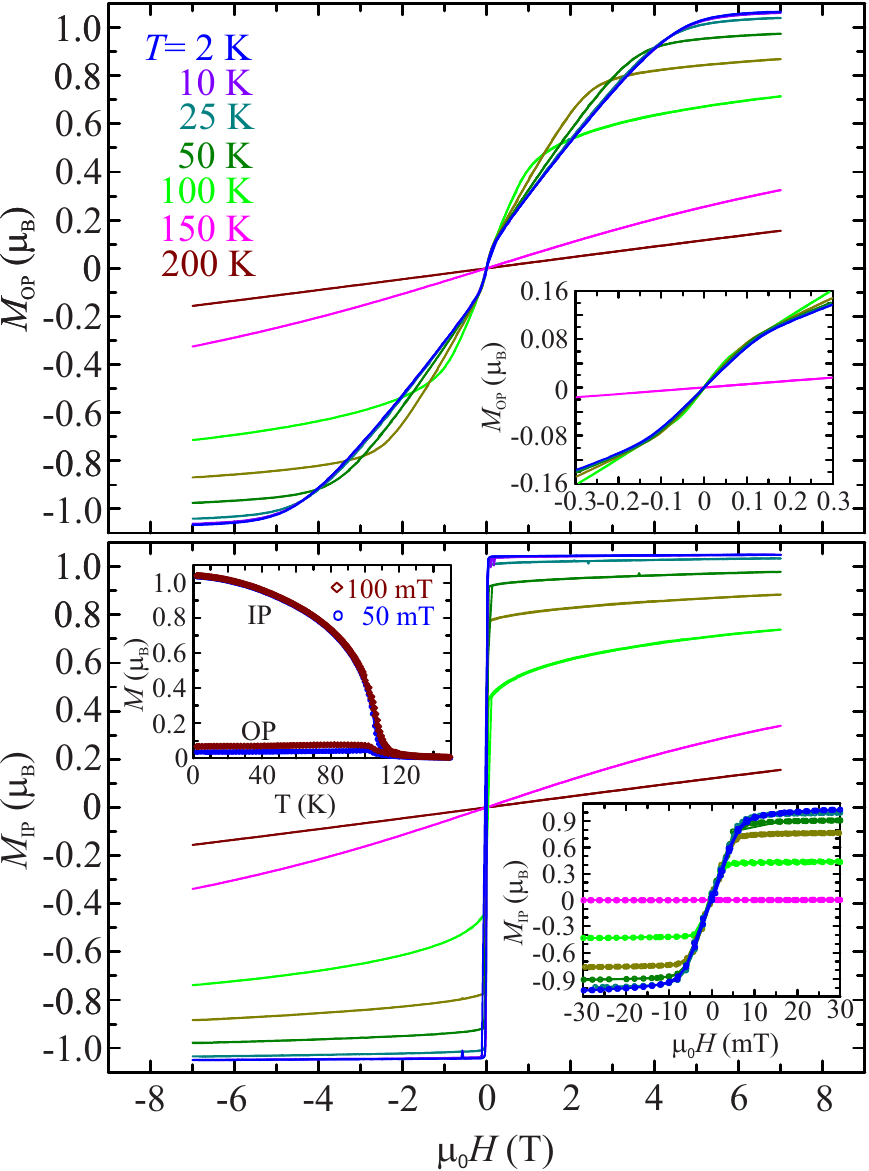}\;
\caption{\label{fig:M-vs-H_expt} Magnetic moment M ($\mu_B/$f.u.) of
  Mn$_{1/4}$NbS$_2$ as function of the magnetic field oriented
  along [0001] (a) and along [1000] directions. The inset in (a)
  displays the magnetic moment as function of temperature, obtained during
  cooling in low fields of 50 (squares) and 100 mT (diamonds) for both
  the IP and OP field directions.  }   
\end{figure}

The magnetic moment of Mn$_{1/4}$NbS$_2$ as function
of the magnetic field $M(H)$ ($\mu_B/$f.u.) is shown in Fig.\
\ref{fig:M-vs-H_expt} for different 
temperatures. All $M(H)$ curves do not exhibit detectable signature of a field
hysteresis. The shape of the magnetization curves shown in Fig.\
\ref{fig:M-vs-H_expt} evidence that Mn$_{1/4}$NbS$_2$ is a soft
easy-plane ferromagnet. For both orientations of the magnetic field,
in-plane (IP) and out-of-plane (OP), a saturation magnetization at $T
= 2.0$~K of $1.05 \mu_B$/f.u. is 
attained, indicating a local magnetic moment $M_{Mn} = 4.2  \mu_B$ per
Mn ion. The observed behavior and the derived characteristic values are
in fair agreement with a previous investigation 
by Onuki et al. \cite{OIHK86}. 
Magnetization data measured while cooling or warming 
in low magnetic fields indicate a ferromagnetic (FM) ordering at
$T_C = 104(2)$~K. The same critical temperature is obtained from the
heat capacity measurements (SM, Fig. 1). 

\begin{figure}[h]
\includegraphics[width=0.4\textwidth,angle=0,clip]{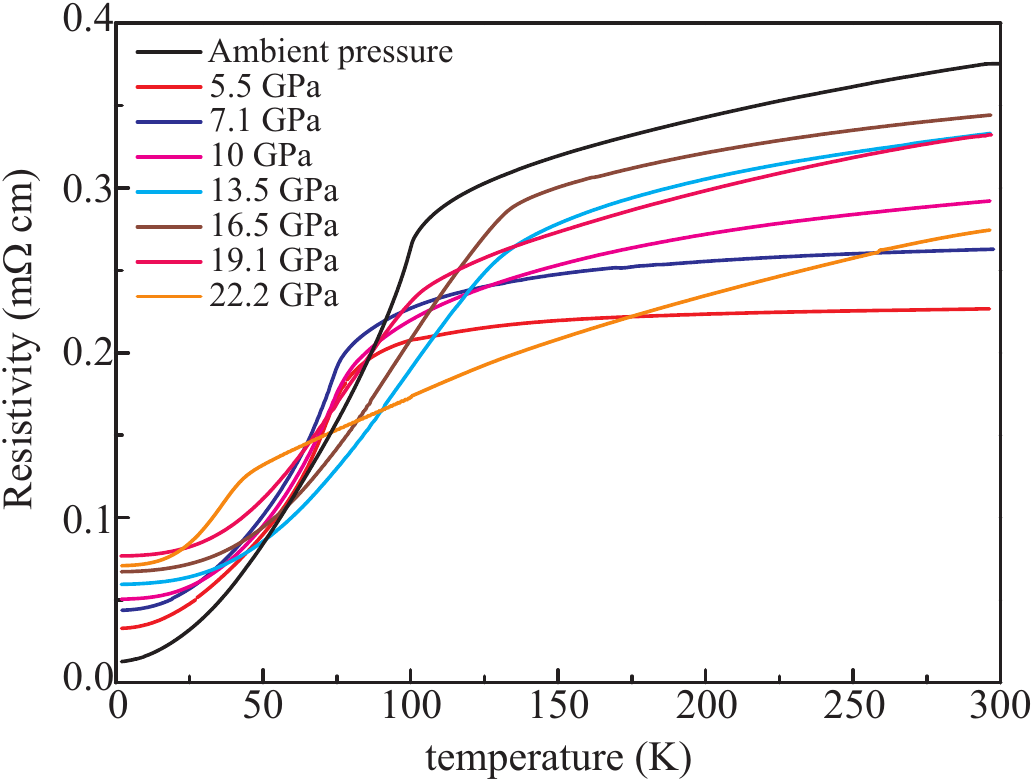}\;
\caption{\label{fig:Resist_expt} Electrical resistivity obtained experimentally
  for  Mn$_{1/4}$NbS$_2$ at zero magnetic field,  plotted as function of temperature for
  different pressure values. }   
\end{figure}

%
The electrical resistivity $\rho(T)$  of Mn$_{1/4}$NbS$_2$ (see Fig.\
\ref{fig:Resist_expt}) measured at different pressure  exhibits well
defined metallic-like beviour typical for magnetically ordered systems.
For all pressures it decreases upon sample
cooling down from room temperature, changing the slope at the critical
temperature $T_{c}$ corresponding to a transition to a magnetically
ordered state. At ambient pressure this corresponds to a Curie
temperature $T_C \approx 104$~K as Mn$_{1/4}$NbS$_2$ exhibits FM order
at lower temperatures.  A pressure 
increase up to 10 GPa results in a continuous decrease of the critical
temperature down to $\sim 75$ K at 10 GPa.
However, it increases abruptly to
$\sim 135$ K at 13.5 GPa, while a further pressure increase results in a
decrease of the critical temperature reaching 40 K at a pressure of 22.2
GPa; the highest pressure attained in this study. 

\begin{figure}[h]
\includegraphics[width=0.5\textwidth,angle=0,clip]{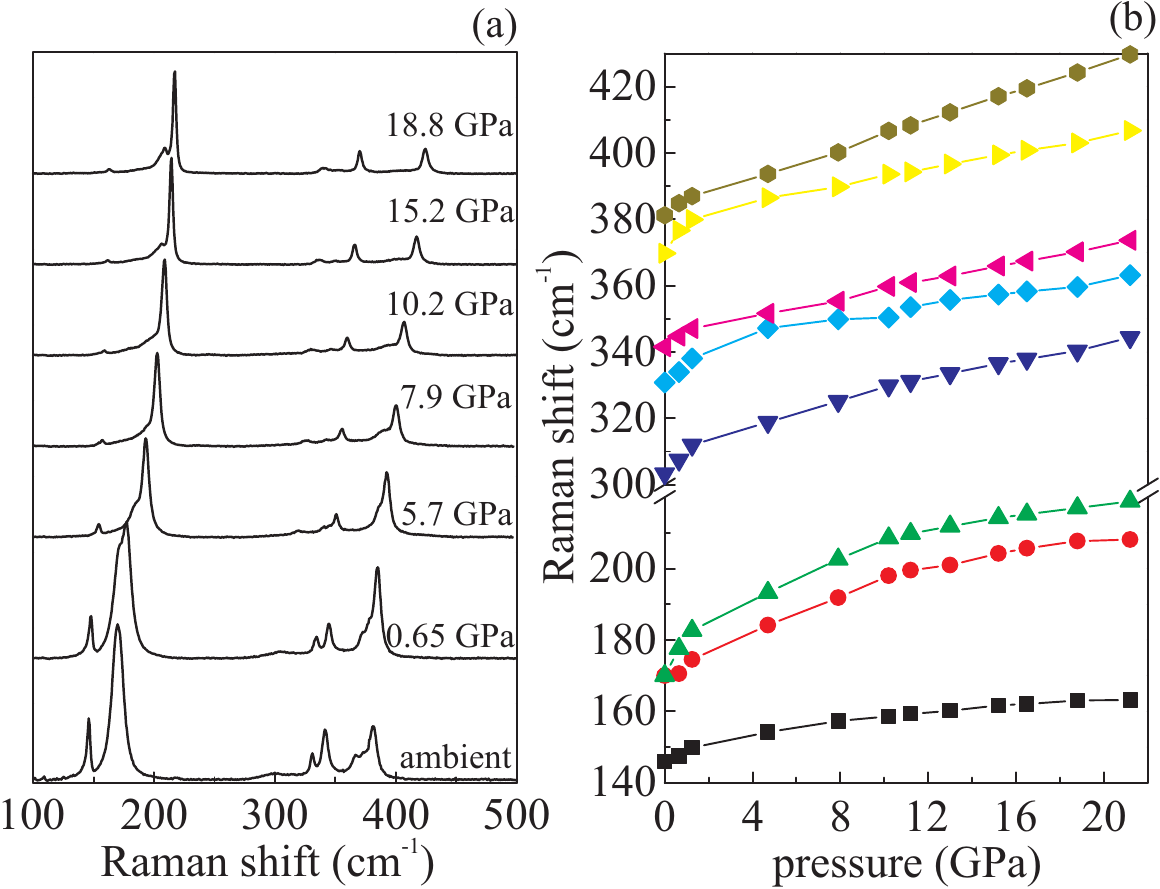}\;
\caption{\label{fig:Raman_expt} Pressure evolution of Raman spectra (a)
  and pressure dependence of the frequencies of the observed Raman peaks
  (b) }   
\end{figure}

The sudden increase of the critical temperature at 13.5 GPa 
can indicate a change of the crystal structure of
Mn$_{1/4}$NbS$_2$ above 10 GPa.
To monitor the possible structural phase transition, the 
investigation of the Raman spectra changes for Mn$_{1/4}$NbS$_2$ under
pressure have been performed, as is shown in Fig.\
\ref{fig:Raman_expt}(a).
The  Raman spectra at ambient pressure are qualitatively similar to that
of the related isostructural compound Fe$_{0.239}$NbS$_2$
\cite{KWY+00}. All Raman resonances observed at ambient pressure are persistent in
the Raman spectra up to the pressures beyond 20 GPa (Fig.\
\ref{fig:Raman_expt}(a)) while the frequency of these peaks shown in Fig.\
\ref{fig:Raman_expt}(b) shows normal increase with pressure
without any discontinuities indicating no structural phase transition in
this pressure regime. These observations allow to make conclusions about possible
changes of the low-temperature magnetic structure of Mn$_{1/4}$NbS$_2$
above 10 GPa.   
However, we will focus below on the pressure effect in the region up to
$\sim 10$ GPa, while the results obtained at higher pressure need more
detailed investigations.

\begin{figure}[h]
\includegraphics[width=0.4\textwidth,angle=0,clip]{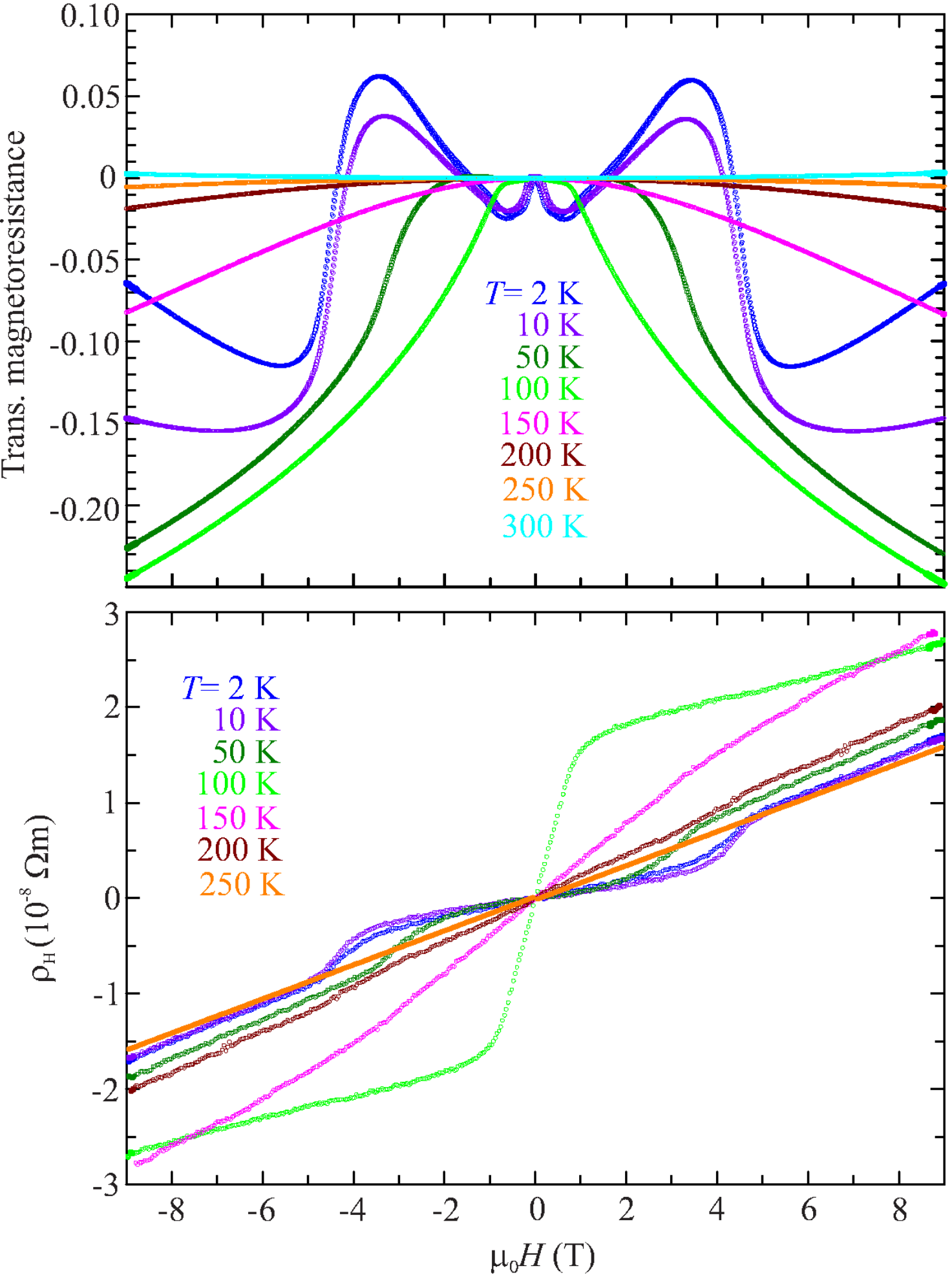}\;
\caption{\label{fig:Hall-MR_expt_p0} Transverse magnetoresistance
  MR$(T,B)$ (top) and Hall resistance $\rho_H(H)$ (bottom) for
  Mn$_{1/4}$NbS$_2$ measured at ambient pressure for different temperatures. The magnetic
  field is applied in the out-of-plane direction.  }   
\end{figure}

%
The transverse MR, which is defined as MR($\%$) = $(\rho(T,H) - \rho(T,0))/
\rho(T,0) \cdot 100\%$, and the Hall resistance $\rho_H(H)$ measured at
ambient pressure are presented in 
Fig. \ref{fig:Hall-MR_expt_p0} as a function of applied field $H$.
The MR for the FM state at $T < 10$~K is positive in the magnetic field
regime $\mu_0H \approx 1.5$ to 3.5 T. This behavior is obviously connected to
the slow turning of the magnetic moments out-of-plane with increasing field,
before reaching magnetic saturation. However, the MR effect is getting
negative at higher magnetic field with the maximum at the saturation point.
The MR is negative for all magnetic fields when the temperature
increases approach $T_C$, showing a maximum at the Curie temperature. 
 The positive slope for the Hall resistivity
(see Fig. \ref{fig:Hall-MR_expt_p0} (b)) indicates a dominating hole-type
conductivity in Mn$_{1/4}$NbS$_2$ at ambient pressure. 
 While linear at high temperatures, the Hall resistivity $\rho_{H}(H)$
 below $T_C$ has a contribution due to the anomalous Hall
 effect vanishing in the magnetically disordered state.
 
\begin{figure}[h]
\includegraphics[width=0.4\textwidth,angle=0,clip]{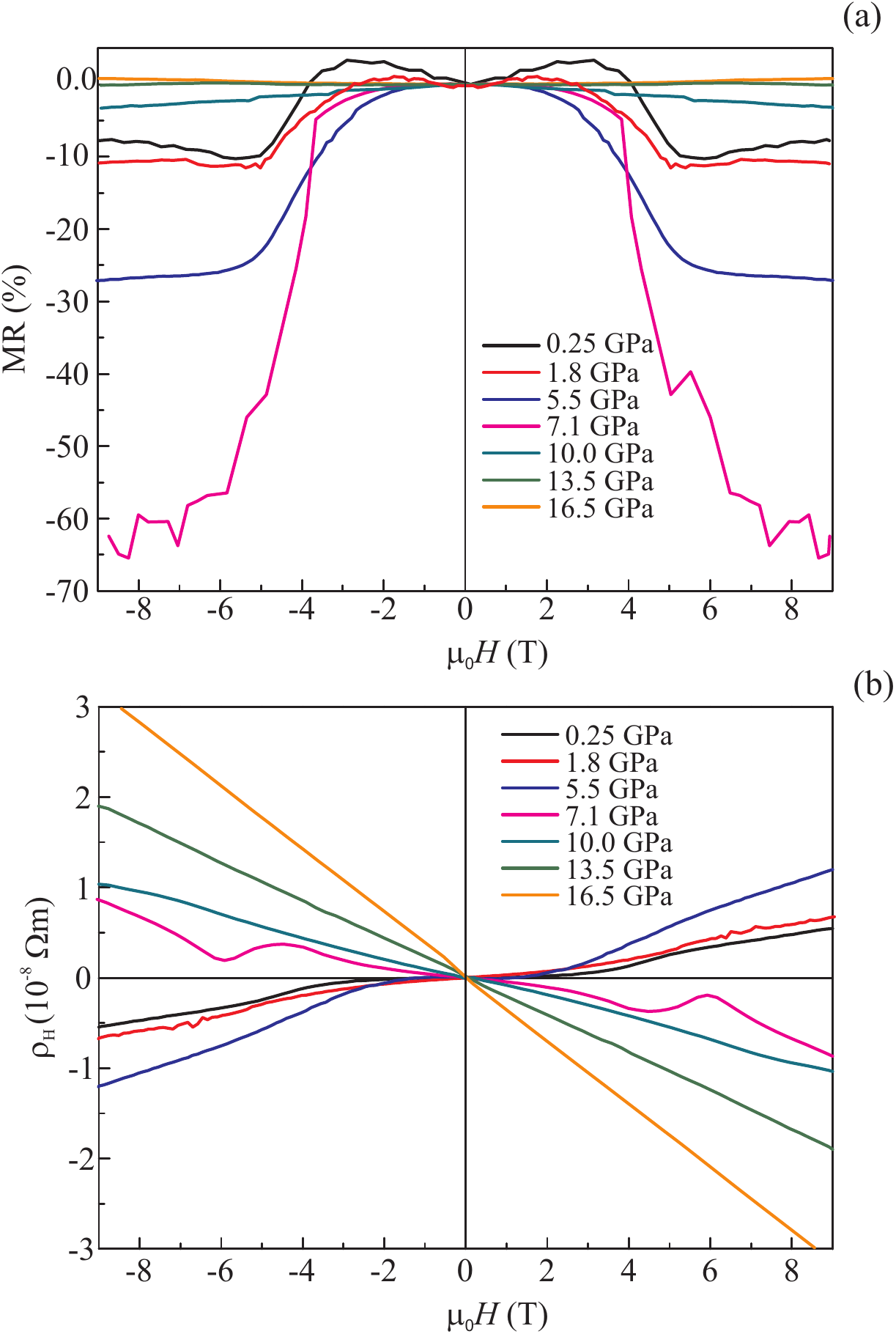}\;
\caption{\label{fig:Hall-MR_expt} Pressure and temperature dependent
  transverse magnetoresistance (a) and Hall resistivity (b) at T = 2K
  for Mn$_{1/4}$NbS$_2$ plotted as function of  external magnetic field. }   
\end{figure}

The magnetoresistance and Hall resistivity measured in the DAC at the
lowest applied pressure 0.25 GPa (SM, Fig. 2 (b))
demonstrate qualitatively a similar behaviour compared to  ambient 
pressure. However, application of a higher pressure has a significant impact
on the shape of the Hall resistance $\rho_H(H)$ and magnetoresistance
MR$(H)$. The low-temperature magnetotransport data (Fig.\
\ref{fig:Hall-MR_expt}, as well as Figs. 2 (b), (d), (f), (h) in SM) indicate the
pressure-induced change of the MR that 
can be associated with the modification of magnetic properties of
Mn$_{1/4}$NbS$_2$. The positive MR observed at ambient pressure is
continuously suppressed while the absolute value of the negative MR at 
the highest fields continuously increases (Fig.\ \ref{fig:Hall-MR_expt}
(a)). For $p = 5.5$ GPa, MR is negative for all applied
magnetic fields, for all temperature regimes (SM, Fig. 2 (d)).
Approaching the pressure 7.1 GPa, MR in the  high filed regime increases 
up to 65\% (Fig.\ \ref{fig:Hall-MR_expt}(a)). At $p = 10$ GPa, MR
drastically drops to 3.2\% at 9 T although it remains negative with nearly
parabolic dependence. Above 10 GPa, MR suddenly decreases to the very
low positive values (only about 0.1\% at 9 T at 13.5 GPa).


Hall resistivity curves $\rho_H(H)$ at pressures up to 5.5 GPa at the
low temperature (Fig.\ \ref{fig:Hall-MR_expt} (b); SM, Fig. 2 (a), (c),
(e), (g))  are qualitatively 
similar to that for ambient pressure demonstrating two regimes.  Two
regimes in the Hall resistivity curves are still persistent at further
pressure increase as can be seen from the $\rho_H(H)$ curve at 7.1  GPa
(Fig.\ \ref{fig:Hall-MR_expt}(b)). At $p = 10$ GPa, Hall resistivity
$\rho_H(H)$ is sublinear, while above 10 GPa  $\rho_H(H)$  is linear at
all temperatures (above and below the critical temperature derived from
the temperature dependence of resistivity) with a negative slope
revealing dominating electron conductivity in Mn$_{1/4}$NbS$_2$ and $\rho_H(H)$
indicating the suppression of ferromagnetic order.
For the magnetically ordered state ($T < T_c$) of  Mn$_{1/4}$NbS$_2$,
when the temperature increases, one can clearly see the contribution due to
the extraordinary Hall effect vanishing in the magnetically disordered
state (SM, Fig.\ 2 (a), (c), (e) and (g)). As it follows from the
results for different pressure values $\rho_{AHE}$ changes the sign twice
upon pressure increase. 

\subsection{Theory}

To allow a detailed interpretation of the experimental results, theoretical
investigations based on first-principles electronic structure
calculations have been performed by using the SPR-KKR band structure
method \cite{SPR-KKR7.7,EKM11}. The DFT calculations based on local density
approximation (LSDA) for FM Mn$_{1/4}$NbS$_2$ at ambient pressure lead
to an underestimation of 
the Mn spin magnetic moment ($3.3 \mu_B$ vs. experimental $4.2 \mu_B$ per
atom), and to a suppression of the Mn spin magnetic moment at high
pressure.  
Therefore, the calculations have been extended treating
correlation effects beyond LSDA, using the LSDA+U approach with the
Hubbard $U$ parameter for Mn of $U = 3$ eV. This leads to an increase of
the spin magnetic moment of Mn at ambient pressure up
to $3.8 \mu_B$ as well as to a stabilization of the finite Mn spin magnetic
moment on the Mn atom when the pressure increases up to 11 PGa.

The pressure-volume dependence, $p-V$,  determined for ordered
Mn$_{1/4}$NbS$_2$ using the VASP package \cite{KH93,KH94} (see SM, Fig. 3),  
shows a linear $p-V$ dependence up to a pressure of $\sim 9$ GPa. At
higher pressures one can expect an instability of the 
original crystalline or magnetic phase. This is in line with the
experimental results that show significant changes of the MR when the
pressure increases above 10 GPa. On the other hand, the experimental
Raman spectra do not exhibit any evidence for a structural phase transition in this
pressure regime.
As it follows from the electronic structure calculations within the
LSDA+U approach, a pressure increase up to  $p=8$ GPa leads to a
decrease of the Mn spin magnetic moment from $m = 3.8 \mu_B$ at ambient
pressure down to $m = 3.2 \mu_B$ at $p = 8$ GPa and to $2.21 \mu_B$ at
10.5 GPa. Thus, one may speculate about a possible pressure induced
transition to the low-spin magnetic state. However, more detailed
investigations  on the properties of Mn$_{1/4}$NbS$_2$, both theoretical
and experimental, are required for this pressure regime, while we will
focus here on the low-pressure phase exhibiting linear $p-V$
dependence shown in Fig.\ 3 (SM).  

\begin{figure}[h]
\includegraphics[width=0.4\textwidth,angle=0,clip]{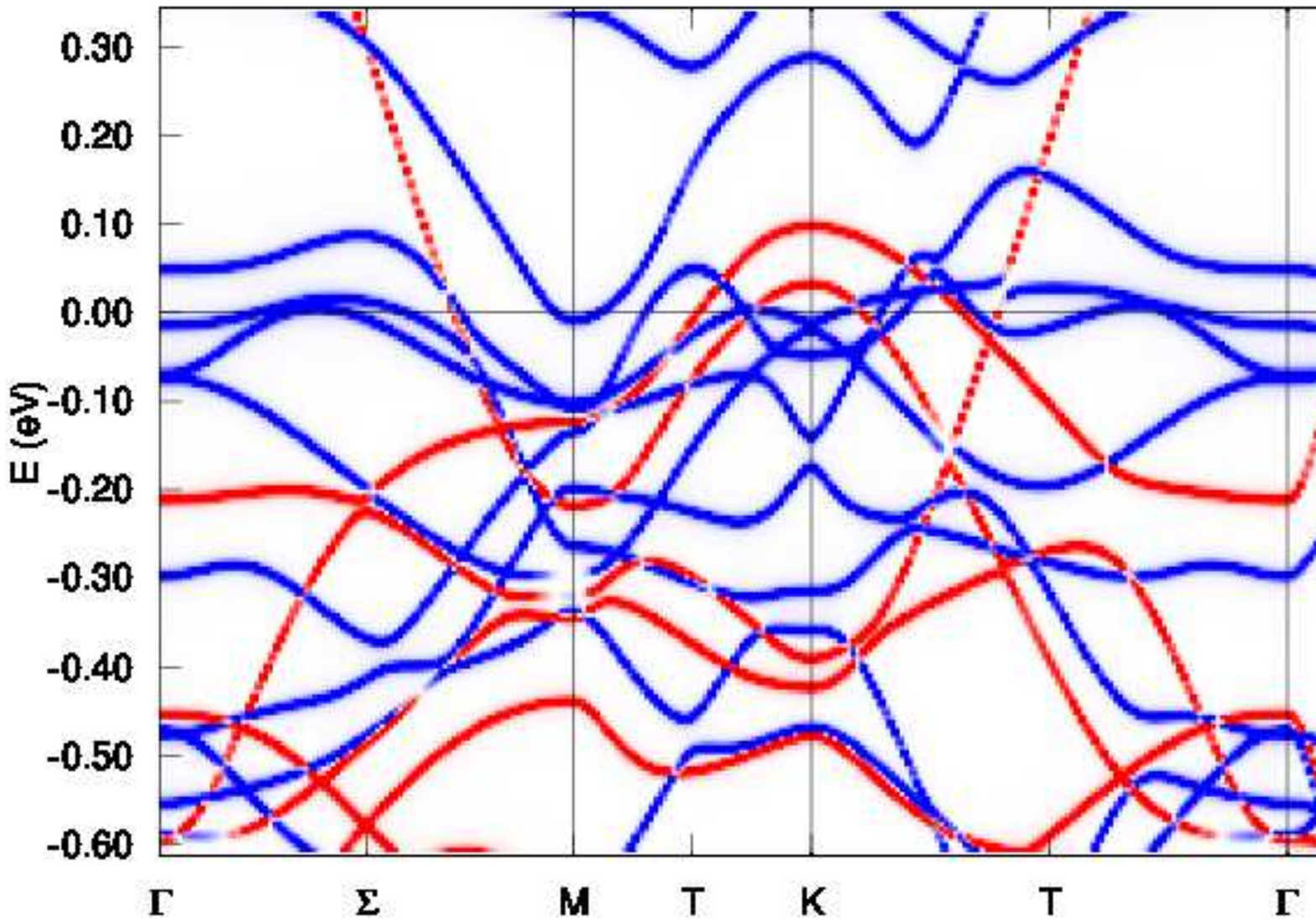}\;(a)
\includegraphics[width=0.4\textwidth,angle=0,clip]{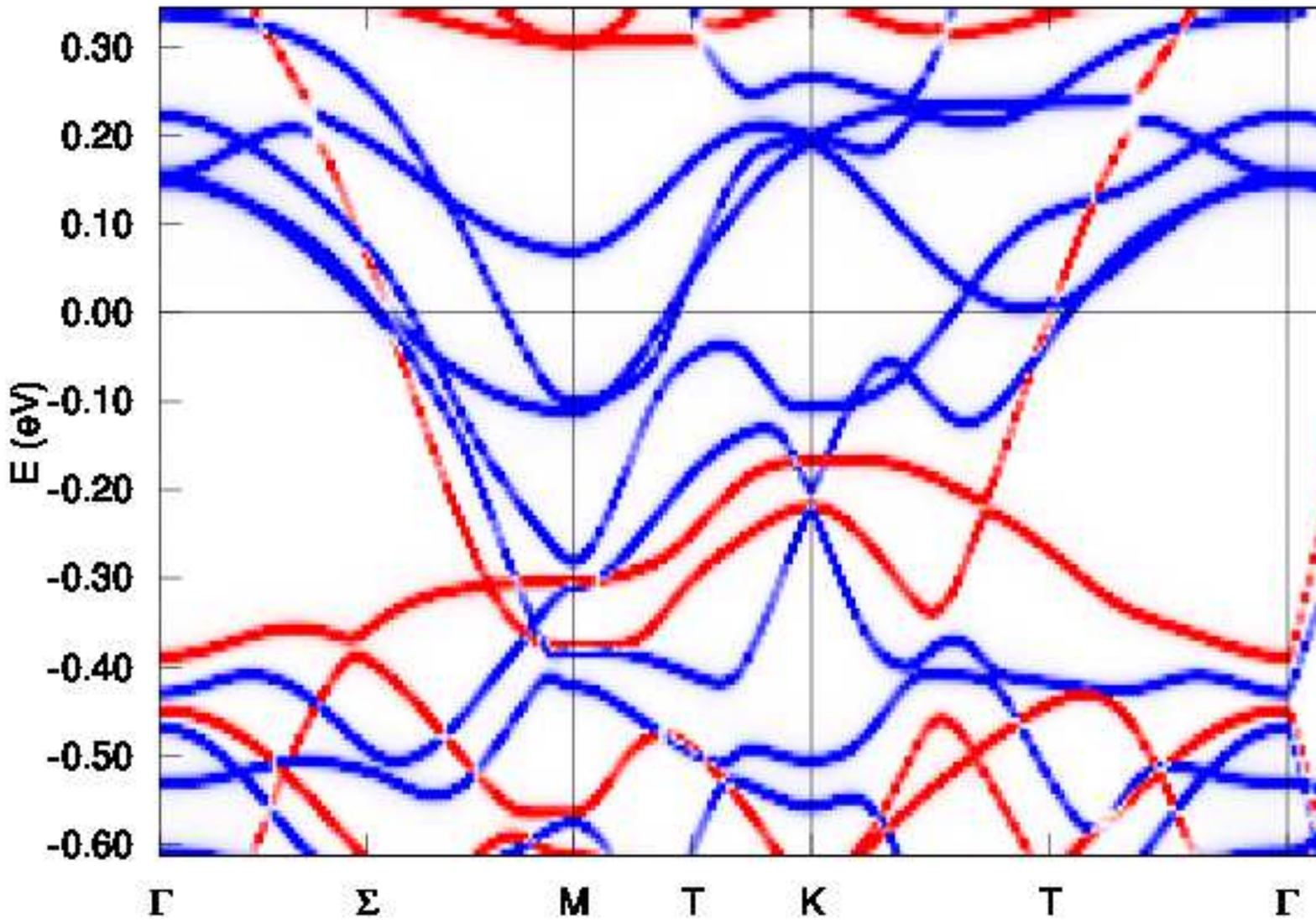}\;(b) \\
\includegraphics[width=0.10\textwidth,angle=270,clip]{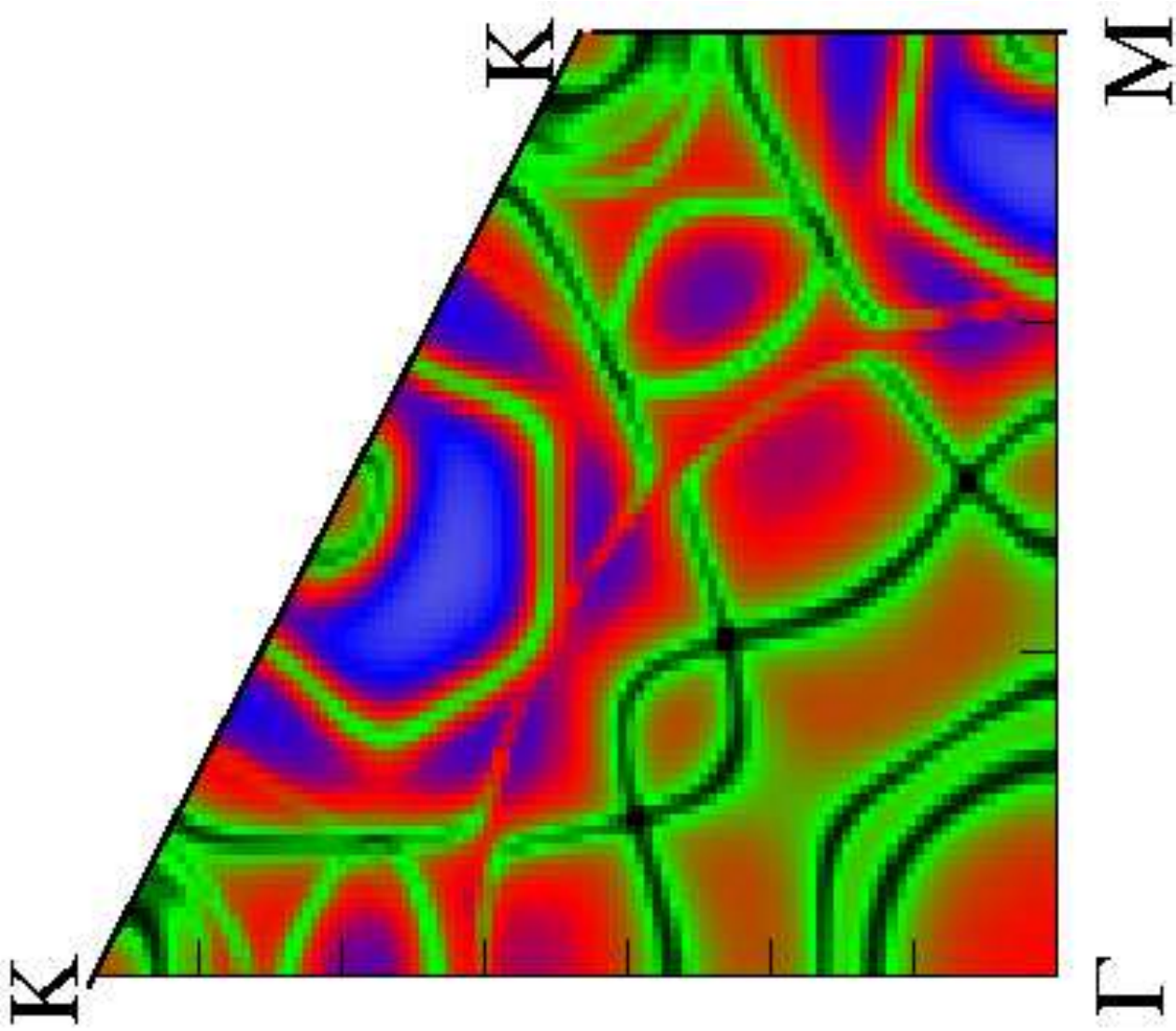}\; 
\includegraphics[width=0.10\textwidth,angle=270,clip]{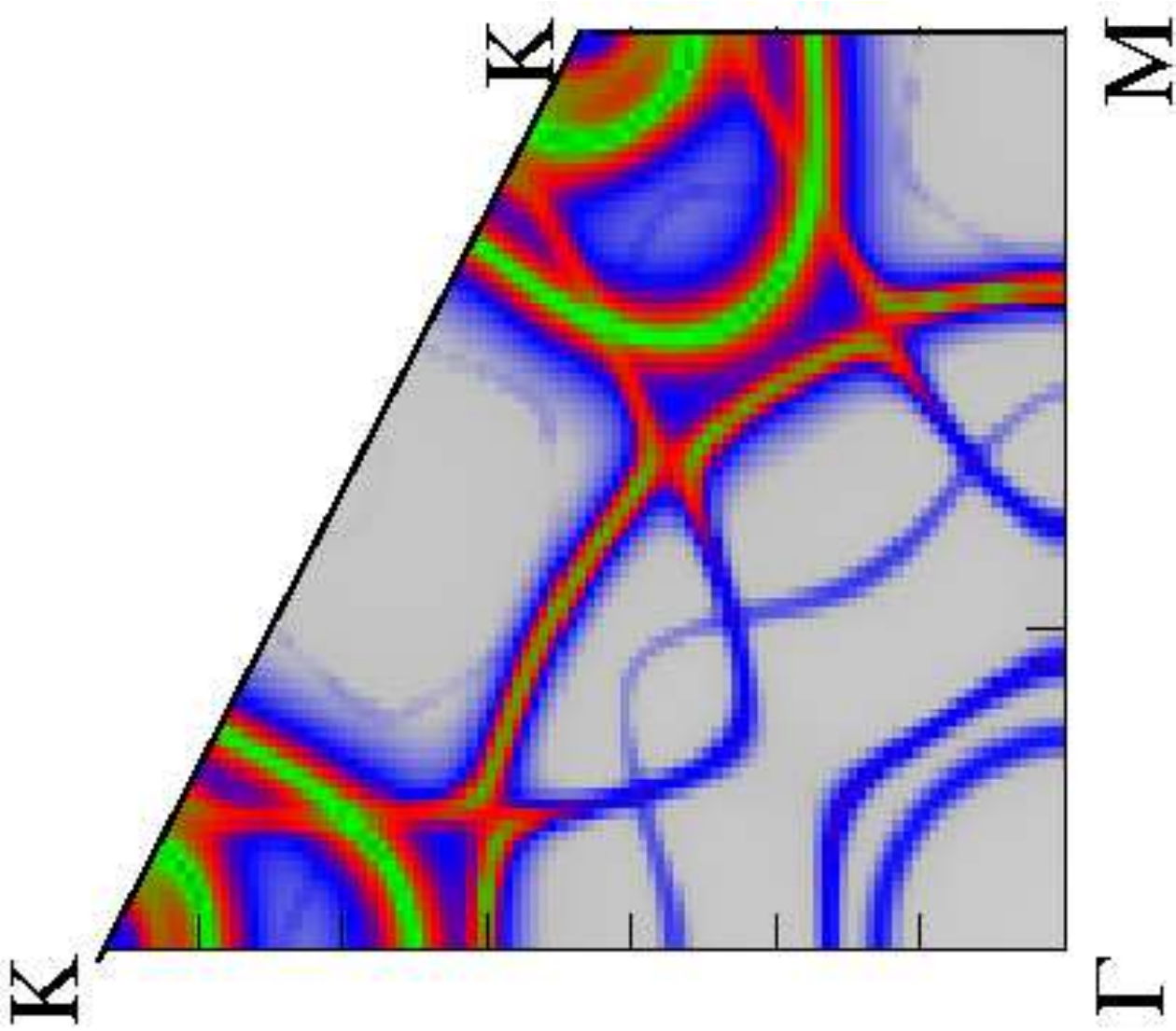}\; (c)
\includegraphics[width=0.10\textwidth,angle=270,clip]{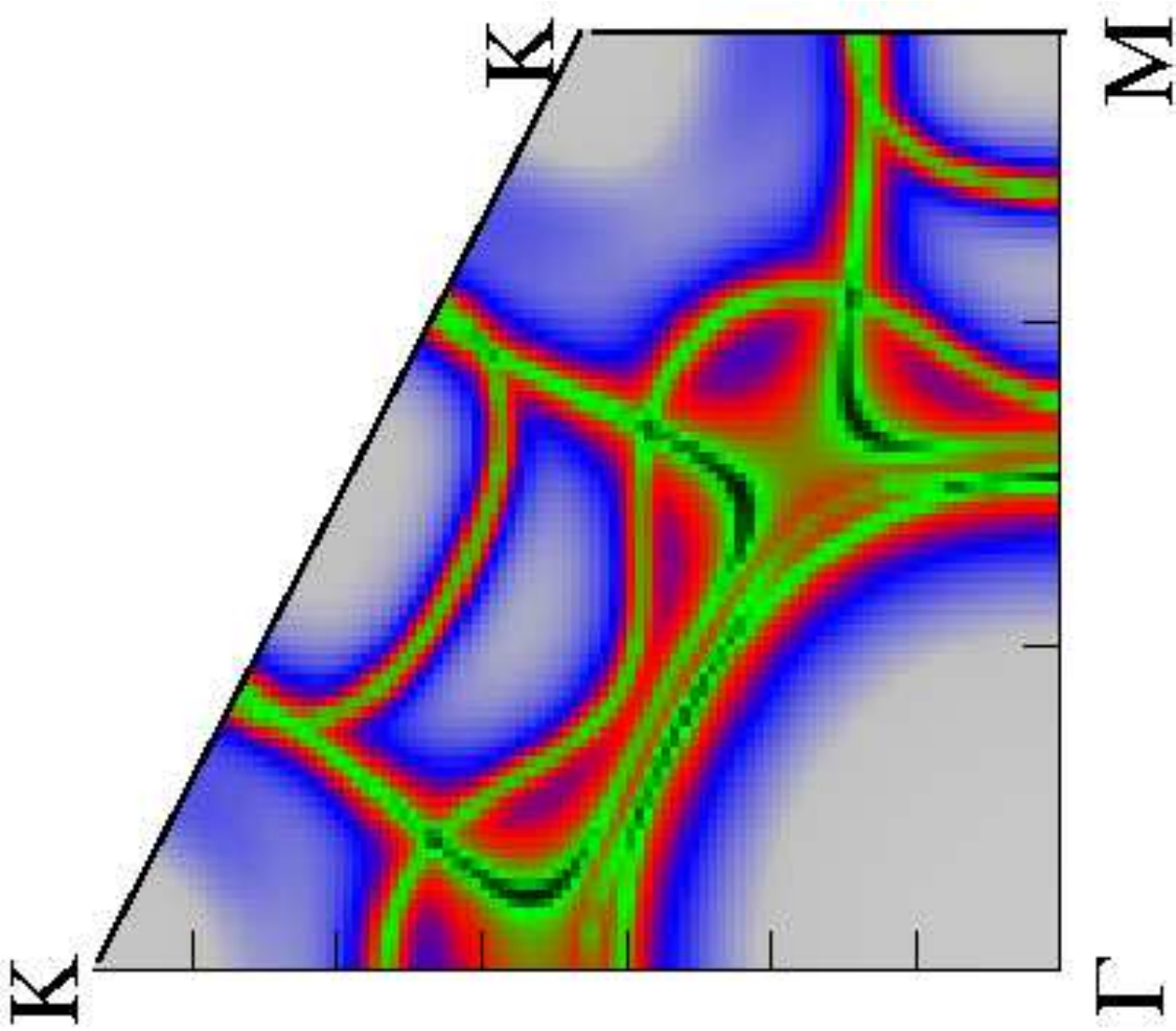}\; 
\includegraphics[width=0.10\textwidth,angle=270,clip]{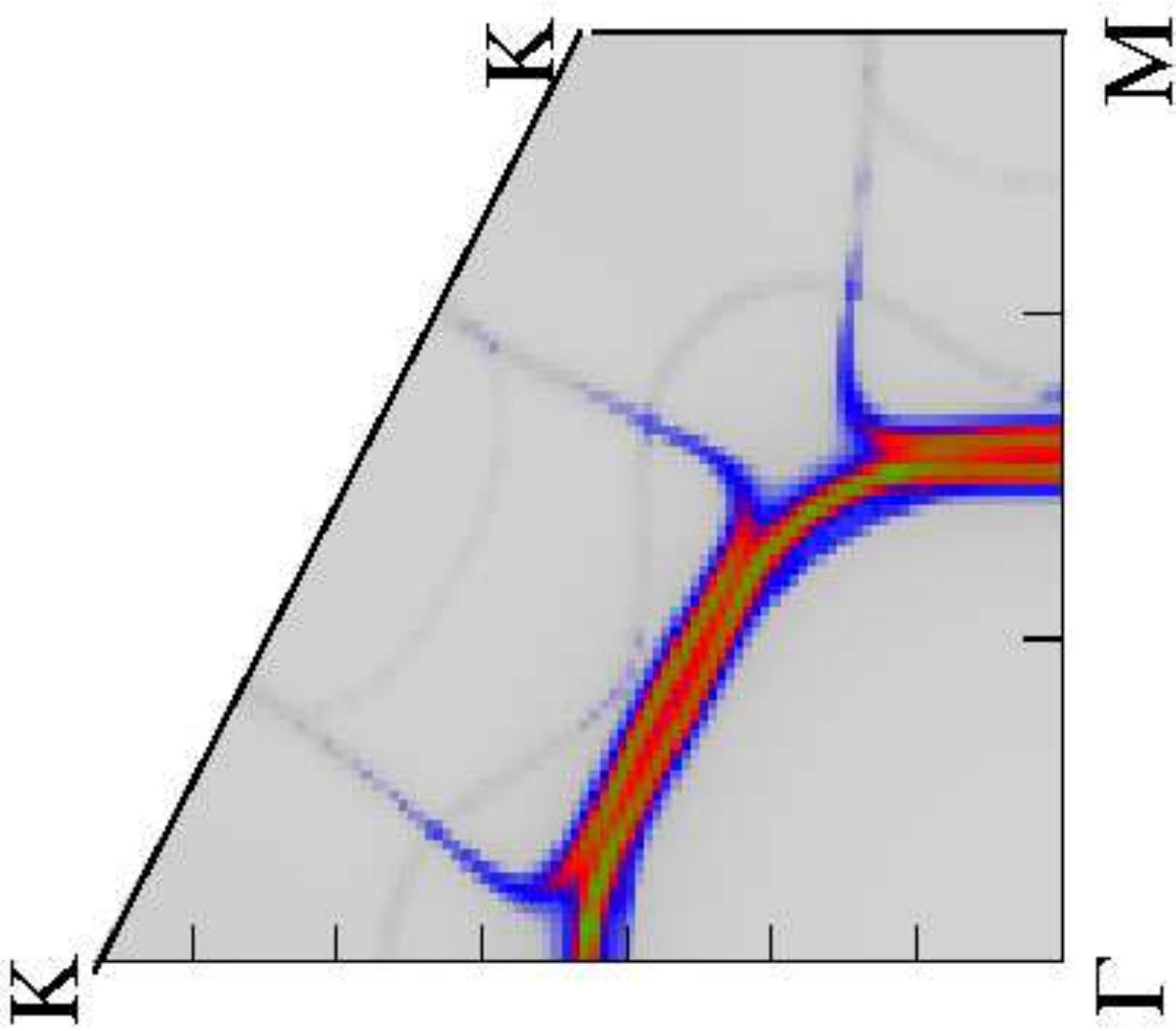}\; (d)
\caption{\label{fig:BSF_pressure} Bloch spectral function for
  Mn$_{1/4}$NbS$_2$ under pressure for $p = 0$ and  $8.0$~GPa, (a)
  and (b), respectively. Blue symbols display majority-spin states
  and red symbols - minority-spin states. (c) and (d) represent the Bloch
  spectral function $A(k_x,k_y,E_F)$ at the Fermi level for Mn$_{1/4}$NbS$_2$ under
  pressure for $p = 0$ and  $8.0$~GPa, respectively. Left panel
  corresponds to majority-spin states, right to minority spin states.   }   
\end{figure}

The properties of the Hall resistivity are determined by the
features of the electronic structure. To demonstrate this,
Fig. \ref{fig:BSF_pressure} shows the calculated Bloch spectral functions
(BSF) $A(\vec{k}, E)$ ((a) and (b)), and $A(\vec{k}_{||},E_F)$ ((c) and
(d)) representing the electronic band structure and Fermi surface
projection onto the (001) plane for the two pressures $p = 0$  and $8$
GPa, respectively. 
A positive slope of the Hall resistivity  $\rho_{H} (B)$ found
experimentally at ambient and small pressures implies a dominating
hole type character of the electric carriers in line with the 
hole-like pockets around the $K$-point of the BZ (see Fig.\
\ref{fig:BSF_pressure} (a)) created by the unoccupied
top of the minority-spin bands (red symbols). This is also seen in the
Fermi surface cut shown in Fig.\ \ref{fig:BSF_pressure} (c) (right
panel) representing the Fermi 
surface created by the  minority-spin states, 
as well as the hole-like pockets created by the majority-spin states 
(\ref{fig:BSF_pressure} (a) blue symbols) along the $\Gamma-M$ and $K-M$ symmetry
directions.
When the pressure increases above $3.5$~ GPa, the minority-spin pockets 
at the $K$ point disappear as the corresponding energy bands (red) move down
in energy, as can be seen in Fig. \ \ref{fig:BSF_pressure} (b) for $p =
8.0$ GPa.
At this pressure, the shape of the Fermi surface has
essentially electron-like features (see Fig.\
\ref{fig:BSF_pressure}(d)), leading to  
the electron-like character for the ordinary Hall effect,
although the hole-like pockets created by the majority-spin states
around $\Gamma$ point still survive also at this pressure.

The MR is a characteristic of the change of the
electrical resistivity $\rho(T,B)$ in the presence 
of an external magnetic field $B$. The corresponding resistivities
$\rho(T,B)$ are calculated for the mono-domain system,
as a function of the temperature. Therefore, in the case of FM-ordered
Mn$_{1/4}$NbS$_2$, these results should be compared with the
experimental data for an applied magnetic field strong 
enough to ensure magnetization saturation in the system. 

\begin{figure}[t]
\includegraphics[width=0.3\textwidth,angle=0,clip]{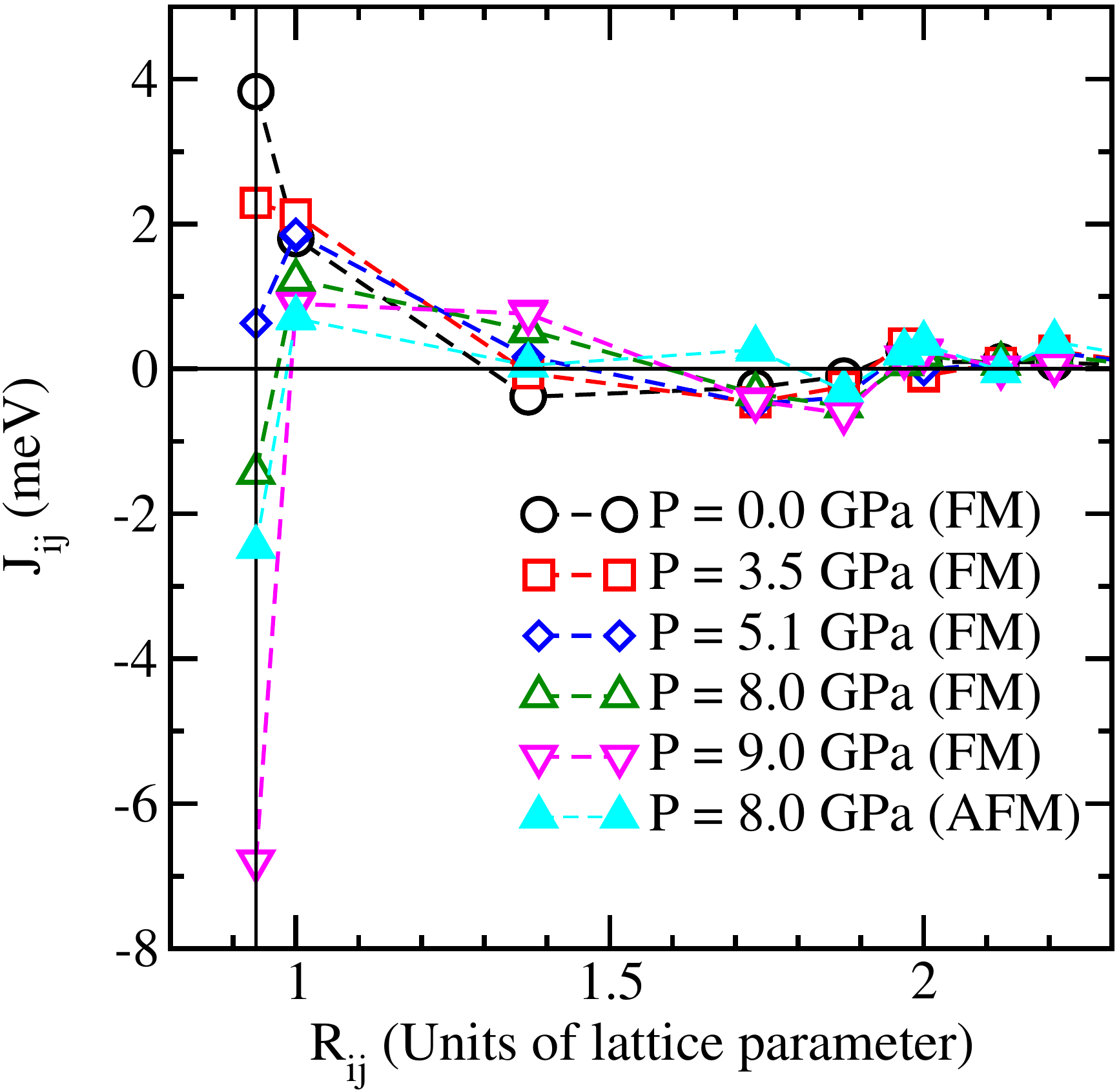}\;(a) 
\includegraphics[width=0.3\textwidth,angle=0,clip]{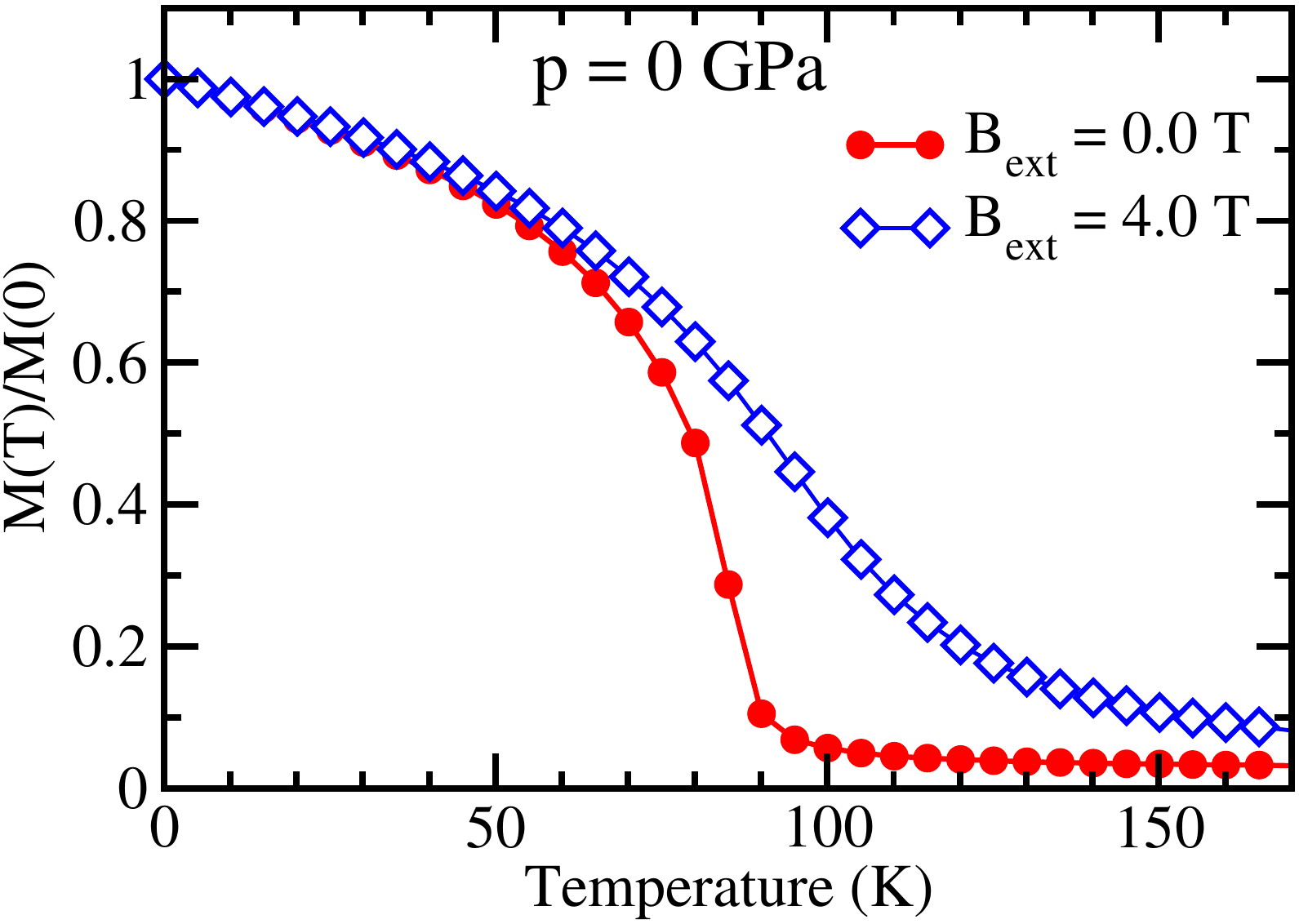}\;(b)
\includegraphics[width=0.3\textwidth,angle=0,clip]{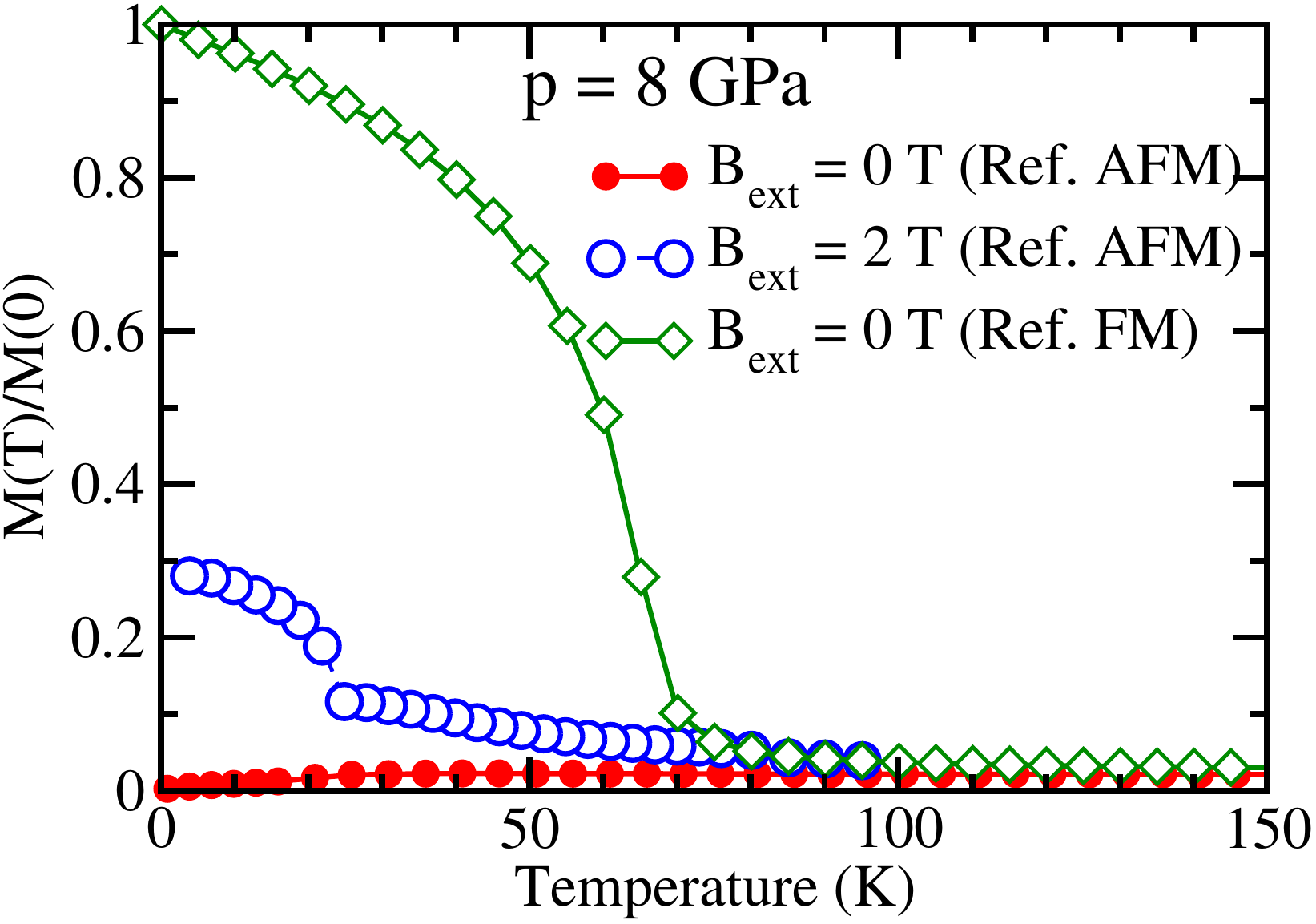}\;(c)
\includegraphics[width=0.3\textwidth,angle=0,clip]{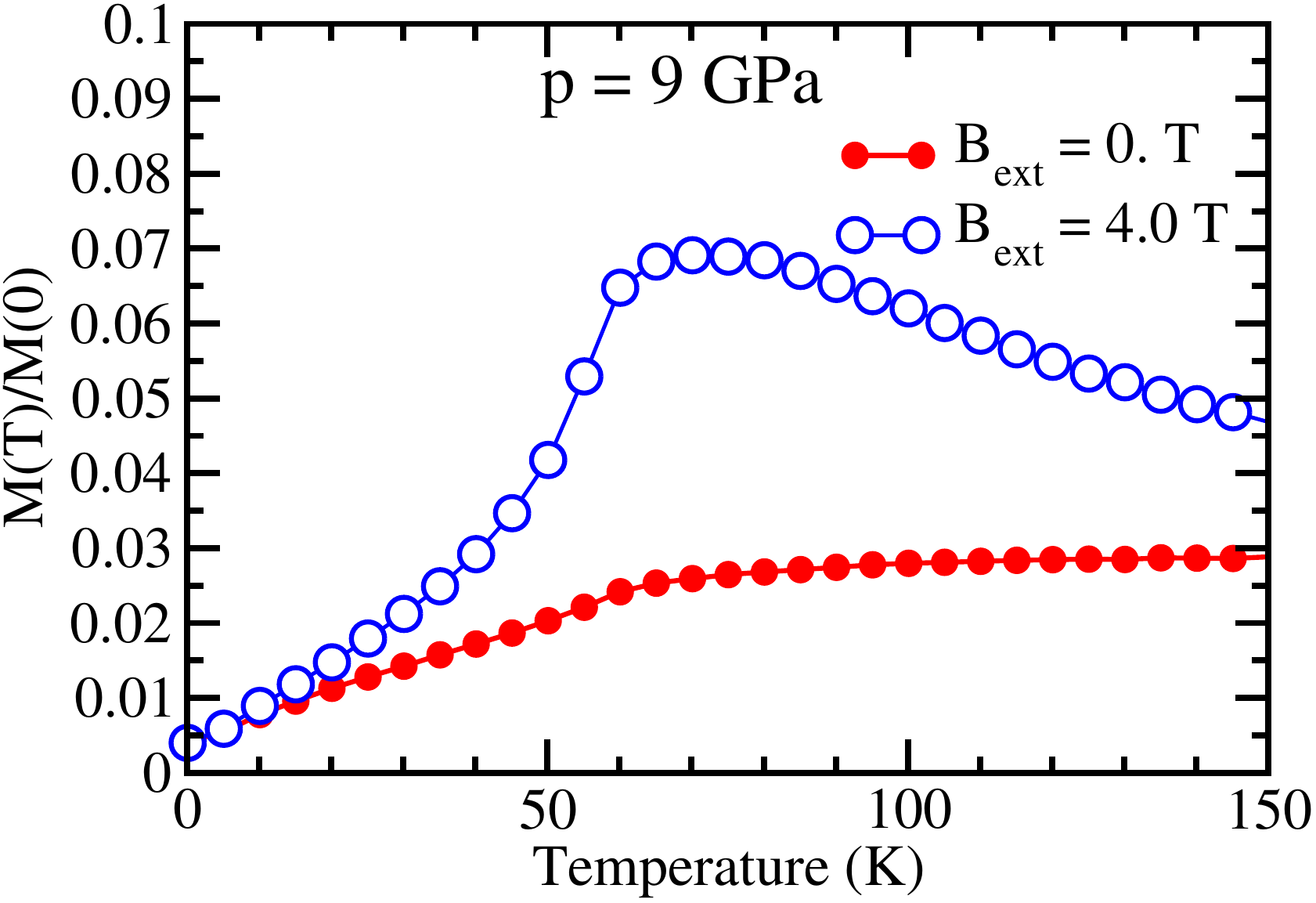}\;(d)
\caption{\label{fig:MC_M-T} (a) Pressure dependent exchange interactions
  for Mn$_{1/4}$NbS$_2$; Open symbols represent the results calculated
  for the FM reference states, closed symbols the AFM reference state.
  The MC results for the temperature dependent magnetization $M(T)$ of 
  Mn$_{1/4}$NbS$_2$ for different pressure:
  $M(T)$ calculated for $B = 0$ (closed circles) and 
  $4.0$~T (open circles) at $p = 0.0$ (b),  $p = 8.0$ (c) and $p = 9.0$ GPa (d).  }   
\end{figure}

\begin{figure}[t]
\includegraphics[width=0.4\textwidth,angle=0,clip]{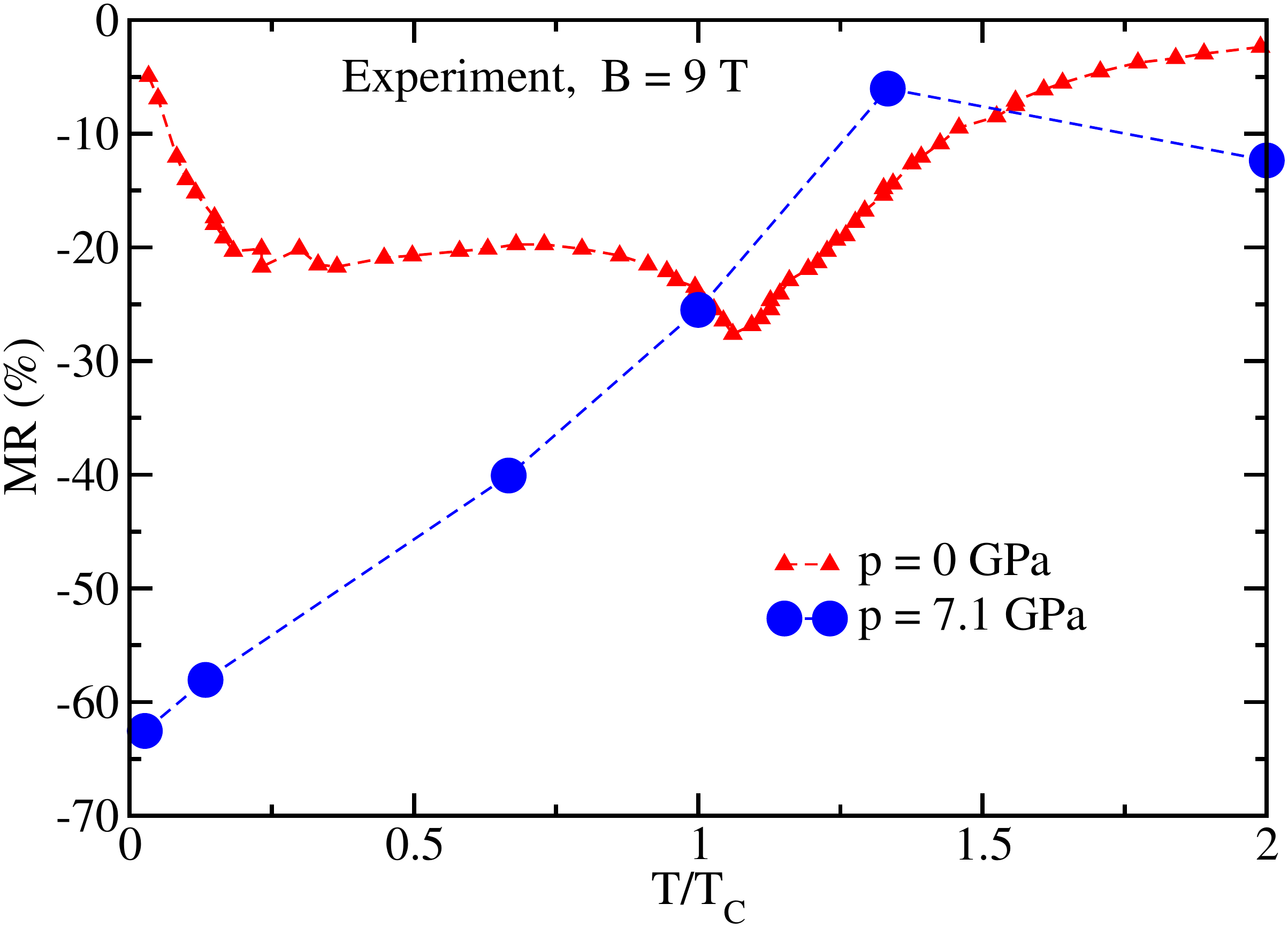}\; (a)\\
\includegraphics[width=0.4\textwidth,angle=0,clip]{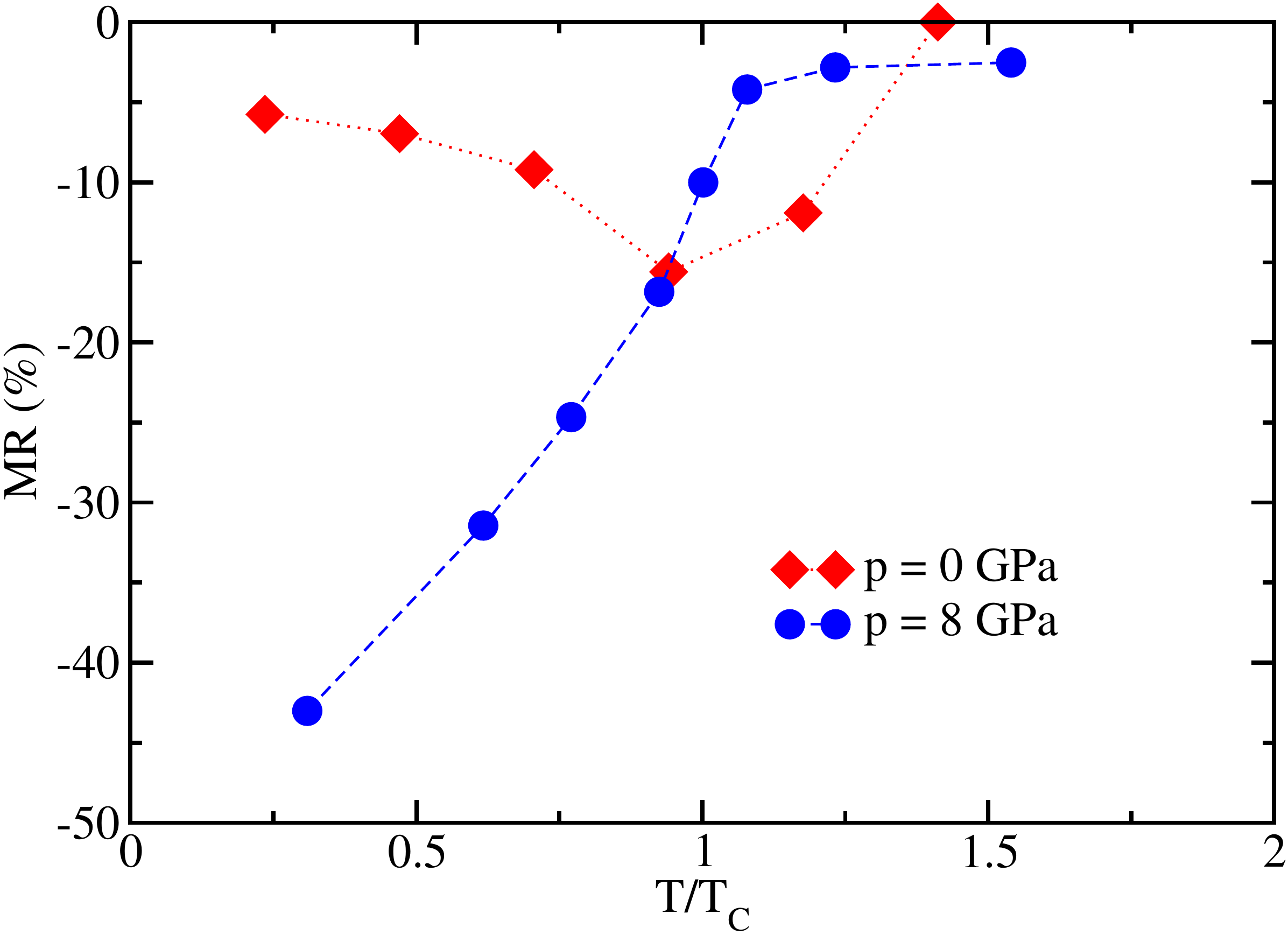}\; (b)
\caption{\label{fig:Magneto-Resist} (a) The temperature dependent
  magnetoresistance of  Mn$_{1/4}$NbS$_2$: (a) Experimental results for
  the magnetic field $B = 7.5$ T and the two pressures 0 and 7.1 GPa;
  (b) theoretical resultes for the magnetic field $B = 4$~T used
  in the case of $p = 0$ and 5 GPa. The results for $p = 8$ GPa are
  obtained using the resistivities for the FM and AFM states, MR =
  $(\rho(T,{\rm FM}) - \rho(T,{\rm AFM}))/ \rho(T,{\rm AFM})$. }   
\end{figure}

The temperature dependent electrical resistivity of magnetically ordered 
metallic systems is determined by thermally induced lattice vibrations
and spin fluctuations. Because of a weak dependence of
the lattice vibrations on the magnetic field, their contribution to the MR can be
neglected and we will focus on the transverse spin
fluctuations that are strongly affected by the temperature and dependent
on the magnetic structure in the system. 
Monte Carlo (MC) simulations give access to the $M(T,B)$ dependencies for
Mn$_{1/4}$NbS$_2$ characterizing the amplitude of the transverse spin
fluctuations and as a result the amplitude of the electron
scattering on thermal spin fluctuations. These simulations use the
exchange coupling parameters obtained within first-principles
calculations performed for the FM reference state. A positive first- and
second-neighbor exchange coupling parameters $J_{ij}$ at low pressure
(see Fig.\ \ref{fig:MC_M-T}(a)) guarantee the FM ground state of the
system. The $M(T, B)$ dependence obtained within MC simulations for
Mn$_{1/4}$NbS$_2$ at ambient pressure is shown in Fig.\ \ref{fig:MC_M-T}
(b) for $B = 0$ T and 4 T, respectively. The resistivities as
a function of the temperature calculated on the basis of $M(T,0)$ and
$M(T,B = 4$ T) lead to the negative magnetoresistance MR$(T)$
 plotted in Fig. \ref{fig:Magneto-Resist} (b). In this case the MR is
governed by the mechanism rather common for FM metals
\cite{YT72}: alignment of the spin magnetic moments  
along the magnetic field reduces the electron scattering
and decreases the resistivity $\rho(T,B = 4$ T) with respect to
$\rho(T,0)$, leading to a negative magnetoresistance in the FM ordered
system \cite{Ued76,MSD+00}.
The impact of the field on the magnetization has a maximum around the
Curie temperature, leading to a maximum of the MR in this temperature
region. This is in good agreement with the experimental findings as it
is shown in Fig. \ref{fig:Magneto-Resist} (a).

When the pressure increases, the calculated exchange coupling parameters $J_{ij}$
exhibit significant changes, as shown in Fig.\ \ref{fig:MC_M-T}(a).
The first-neighbor interaction parameters corresponding to an interaction
between the Mn ions located in neighboring Mn layers in 
Mn$_{1/4}$NbS$_2$ become negative when the pressure is approaching $p = 8$ GPa.
This should lead to an antiferromagnetic alignment of the magnetic
moments of these layers if no other interactions are taken into
account. The second-neighbor parameters characterizing the Mn-Mn
interactions within the layers are positive, stabilizing the FM order within
the layers, although they decrease with increasing pressure. 
%
The total energy calculations for $p = 8$ GPa give the difference
$E_{FM} - E_{AFM} = 36$ meV per formula unit, indicating stability of
the AFM state.
The dependence of the exchange coupling parameters on the magnetic  
configuration can be crucial to describe metamagnetic phase
transitions  properly as it was found for instance discussing the AFM-FM
transition in FeRh \cite{PMK+16}.
%
%
%
Therefore, the calculations of $J_{ij}$ for Mn$_{1/4}$NbS$_2$ at $p=8$
GPa have been performed for the AFM reference state with the
layer-by-layer antiferromagmetic alignment of 
the magnetic moments. The resulting Mn-Mn exchange coupling parameters 
presented in  Fig. \ref{fig:MC_M-T}(a) by closed triangles, stabilize
the AFM state in the system. This is shown by MC simulations taking into
account the exchange interactions only within a sphere with radius $R = 2a$,
with $a$ the lattice parameter (see
Fig. \ref{fig:MC_M-T}(c), closed circles). The external magnetic field  
pushes the magnetic moments towards the FM alignment (see open circles in
Fig. \ref{fig:MC_M-T} (c)). This leads to a
modification of the Mn-Mn exchange interaction parameters
(Fig. \ref{fig:MC_M-T}(a), open triangles) and to a
stabilization of the FM state. Note that in the case of ambient
pressure, the exchange parameters calculated both for the FM as well as 
for the AFM reference state indicate stability of the FM state
(Fig. \ref{fig:MC_M-T}(c), diamonds). 
Thus, one can expect a field-induced AFM-FM transition at $p = 8$~GPa.
 MC simulations for the pressure  $p = 9$ GPa  demonstrate the stability
 of the layer-by-layer AFM state (see Fig. \ref{fig:MC_M-T} (d)) for which
the field induced metamagnetic transition is not possible anymore (see
Fig. \ref{fig:MC_M-T} (d), open circles).  

To discuss the behaviour of the magnetoresistance MR($T$) of
Mn$_{1/4}$NbS$_2$ at $p=8$ GPa, the change of the resistivity during the 
AFM-to-FM transition was calculated as a difference of the electrical
resistivities for the AFM (i.e., without magnetic field) and for the FM
(with magnetic field) states of the 
system. The AFM state was approximated by the layer-by-layer
antiferromagnetic structure,  i.e. with two sublattices having
antiparallel alignment of the magnetic moments. For the sake of
simplicity, the temperature dependent magnetization for each sublattice,
$M(T)$, was taken the same as for the FM state calculated for $p = 8$
GPa (open diamonds in Fig. \ref{fig:MC_M-T} (c)).
The resulting MR is shown in Fig. \ref{fig:Magneto-Resist} (b) by
circles as a function of the temperature. A crucial result following from
these calculations is the maximum of the MR at low temperature due to
weak thermal disorder in the system. When the temperature increases the
MR decreases and vanishes at the critical temperature due to a transition
to the paramagnetic state. A similar behavior of the MR has
been obtained experimentally, as it is shown in
Fig. \ref{fig:Magneto-Resist} (a) by circles.

Finally, it is worth to discuss briefly also the pressure dependent
behavior of the non-conventional contribution to the Hall resistivity for
Mn$_{1/4}$NbS$_2$. As it was shown above, the high-field $\rho_{H}(B)$
 extrapolated to the $B = 0$~T limit gives the AH resistivity  
(AHR) assuming FM order in the system, which exhibits a non-monotonous
behaviour changing sign twice upon the pressure increase up to 10 GPa
(see SM, Fig. 2 (a), (c), (e)). However, no indication for a
sign change of the AHR has been found in calculations for the FM state
of the system.  

On the other hand, in the intermediate pressure regime, before the
layer-by-layer AFM state is stabilized, 
a non-collinear AFM structure is expected as a results of
competition between the FM and AFM interatomic exchange interactions.
Moreover, in this case the Dzyaloshinskii-Moriya interactions (DMI)
should have a crucial role for the formation of a chiral magnetic
texture, despite its magnitude ($\sim 0.2$ meV) is smaller by an order of
magnitude when compared to the competing isotropic exchange interactions.
As a result, an additional topological contribution to the Hall
resistivity occurs in the presence of an external magnetic field,
$\rho^{THE}_{xy}$ (topological Hall effect (THE))\cite{NSO+10}
\begin{eqnarray}
  \rho_{xy} &=& \rho^{OHE}_{xy} + \rho^{AHE}_{xy} + \rho^{THE}_{xy}
\end{eqnarray}
with $\rho^{AHE}_{xy} \sim M_{z}$ and  $\rho^{THE}_{xy} \sim
\vec{B}_{eff}$, where $M_{z}$ is the magnetization component along the $z$
direction and $\vec{B}_{eff}$ is the emergent magnetic field having
topological origin and being nonzero in the magnetic textures
characterized by finite scalar chirality.
Note also that a competition of the DMI with the isotropic exchange interactions
and the applied magnetic field can lead also to the formation of more complicated   
topologically nontrivial magnetic textures (e.g. skyrmions) as it 
was predicted recently for Fe$_{1/4}$TaS$_2$ \cite{PMK+16}. 
Both extraordinary contributions to the Hall resistivity increase in the
system with the in-plane magnetic anisotropy due to increase of the external
magnetic field.

\section{Summary}

To summarize, we present the results of experimental and theoretical
investigations on the pressure dependent magnetic and transport
properties of Mn$_{1/4}$NbS$_2$. A strong increase of the
magnetoresistance up to $\sim 60\%$ at low temperature has been
observed experimentally for the pressure $p = 7.1$ GPa. 
To get insight into the driving mechanism behind this phenomenon, theoretical
investigations have been performed based on first-principles
electronic structure calculations combined with Monte Carlo simulations.
As a result, the field-induced metamagnetic AFM-FM transition was
suggested as a mechanism responsible for the high magnetoresistance at
$p = 7.1$ GPa, which is larger than $\sim 50\%$ as was observed for FeRh
\cite{BB95,VLM+13}. This suggests a new family of materials with controllable
metamagnetic transition from AFM to FM ordering that is very promising for
various future applications, i.e. materials based on 
intercalated TMDCs, for which however  tuning of the interatomic
exchange parameters due to composition variation instead of pressure may 
be more appropriate. In particular, one can expect a transition from the
FM to the AFM state upon substitution of Mn by Fe, coming
towards AFM-ordered Fe$_{1/4}$NbS$_2$\cite{GBR+81}.
The pressure induced modification of the Hall resistivity including 
ordinary and extraordinary (anomalous and topological) contributions is
discussed on the basis of calculated electronic structure and exchange
coupling parameters and their modifications induced by increasing
pressure. Mutual analysis of the experimental and theoretical results
allows to make a suggestion about the field-controlled formation of
chiral magnetic structure in the intermediate pressure regime
characterized by strong competition of the FM and AFM interatomic
exchange interactions in  Mn$_{1/4}$NbS$_2$.

\section{Acknowledgments}

The work in Dresden was supported by the German Research Foundation
(DFG) under Projects No. ME 3652/3-1 and GRK 1621 and by the ERC
Advanced Grant No. 742068 "TOPMAT". PGN acknowledges the support by
the Russian Science Foundation (Project No.17-72-20200).
Financial support by the German Research Foundation (DFG) under the
project BE 1653/35-1 and by the state of Schleswig Holstein is
gratefully acknowledged for the work carried out in Kiel. SP and SM
(M\"unchen) acknowledges a financial support from the DFG via SFB 1277 and
via the priority program EB154/36-1. 


%

\end{document}